\documentclass[a4paper,11pt]{article}
\pdfoutput=1 
\usepackage{jheppub} 
\usepackage[T1]{fontenc} 
\usepackage{graphicx}
\graphicspath{{fig/}}
\usepackage{amsmath}
\usepackage{bm}
\usepackage{slashed}
\usepackage{amsfonts}

\allowdisplaybreaks

\newcommand{\state}[4]{{^{#1}\hspace{-0.6mm}{#2}_{#3}^{[#4]}}}
\newcommand{\statenc}[3]{{^{#1}\hspace{-0.6mm}{#2}_{#3}}}
\newcommand\CSaSz{\state{1}{S}{0}{1}}

\newcommand\CScSa{\state{3}{S}{1}{1}}

\newcommand\CScPa{\state{3}{P}{1}{1}}
\newcommand\CScPb{\state{3}{P}{2}{1}}

\newcommand\COaSz{\state{1}{S}{0}{8}}

\newcommand\COcSa{\state{3}{S}{1}{8}}
\newcommand\COcPz{\state{3}{P}{0}{8}}

\newcommand\COcPj{\state{3}{P}{J}{8}}

\newcommand{\initialstate}[2]{{#1^{[#2]}}}
\newcommand\vone{\initialstate{v}{1}}
\newcommand\veight{\initialstate{v}{8}}
\newcommand\aone{\initialstate{a}{1}}
\newcommand\aeight{\initialstate{a}{8}}
\newcommand\tone{\initialstate{t}{1}}
\newcommand\teight{\initialstate{t}{8}}

\newcommand\DeltaZero{\Delta_0}
\newcommand\DeltaPPZero{{\Delta_0''}}
\newcommand\DPlusOne{\Delta_+^{[1]}}
\newcommand\DPlusPOne{{{\Delta_+^{[1]}}'}}
\newcommand\DPlusPPOne{{{\Delta_+^{[1]}}''}}
\newcommand\DPlusEight{\Delta_-^{[8]}}
\newcommand\DPlusPEight{{{\Delta_-^{[8]}}'}}
\newcommand\DPlusPPEight{{{\Delta_-^{[8]}}''}}
\newcommand\DMinusOne{\Delta_-^{[1]}}
\newcommand\DMinusPOne{{{\Delta_-^{[1]}}'}}
\newcommand\DMinusPPOne{{{\Delta_-^{[1]}}''}}
\newcommand\DMinusEight{\Delta_+^{[8]}}
\newcommand\DMinusPEight{{{\Delta_+^{[8]}}'}}
\newcommand\DMinusPPEight{{{\Delta_+^{[8]}}''}}
\newcommand\DPMOne{\Delta_\pm^{[1]}}
\newcommand\DPMEight{\Delta_\pm^{[8]}}
\newcommand\DPMPOne{{{\Delta_\pm^{[1]}}'}}
\newcommand\DPMPEight{{{\Delta_\pm^{[8]}}'}}
\newcommand\DPMPPOne{{{\Delta_\pm^{[1]}}''}}
\newcommand\DPMPPEight{{{\Delta_\pm^{[8]}}''}}

\newcommand\CSaPaT{\state{1}{P}{{1,T}}{1}}
\newcommand\CSaPaL{\state{1}{P}{{1,L}}{1}}
\newcommand\CScSaT{\state{3}{S}{{1,T}}{1}}
\newcommand\CScSaL{\state{3}{S}{{1,L}}{1}}
\newcommand\CScPaT{\state{3}{P}{{1,T}}{1}}
\newcommand\CScPaL{\state{3}{P}{{1,L}}{1}}
\newcommand\CScPbTb{\state{3}{P}{{2,TT}}{1}}
\newcommand\CScPbTa{\state{3}{P}{{2,T}}{1}}
\newcommand\CScPbL{\state{3}{P}{{2,L}}{1}}
\newcommand\COaPaT{\state{1}{P}{{1,T}}{8}}
\newcommand\COaPaL{\state{1}{P}{{1,L}}{8}}
\newcommand\COcSaT{\state{3}{S}{{1,T}}{8}}
\newcommand\COcSaL{\state{3}{S}{{1,L}}{8}}
\newcommand\COcPaT{\state{3}{P}{{1,T}}{8}}
\newcommand\COcPaL{\state{3}{P}{{1,L}}{8}}
\newcommand\COcPbTb{\state{3}{P}{{2,TT}}{8}}
\newcommand\COcPbTa{\state{3}{P}{{2,T}}{8}}
\newcommand\COcPbL{\state{3}{P}{{2,L}}{8}}
\newcommand\CScPL{\state{3,L}{P}{{}}{1}}
\newcommand\COcPL{\state{3,L}{P}{{}}{8}}
\newcommand\CScPT{\state{3,T}{P}{{}}{1}}
\newcommand\COcPT{\state{3,T}{P}{{}}{8}}

\newcommand\CBcPz{\state{3}{P}{{0}}{b}}
\newcommand\CBcPaT{\state{3}{P}{{1,T}}{b}}
\newcommand\CBcPaL{\state{3}{P}{{1,L}}{b}}
\newcommand\CBcPbTb{\state{3}{P}{{2,TT}}{b}}
\newcommand\CBcPbTa{\state{3}{P}{{2,T}}{b}}
\newcommand\CBcPbL{\state{3}{P}{{2,L}}{b}}
\newcommand\CBcPL{\state{3,L}{P}{{}}{b}}
\newcommand\CBcPT{\state{3,T}{P}{{}}{b}}

\newcommand\DAA{{\delta(1-z+\zeta_1)}}
\newcommand\DBB{{\delta(1-z-\zeta_1)}}
\newcommand\DXX{{\delta(1-z+\zeta_2)}}
\newcommand\DYY{{\delta(1-z-\zeta_2)}}
\newcommand\DPAA{{\delta'(1-z+\zeta_1)}}
\newcommand\DPBB{{\delta'(1-z-\zeta_1)}}
\newcommand\DPXX{{\delta'(1-z+\zeta_2)}}
\newcommand\DPYY{{\delta'(1-z-\zeta_2)}}

\newcommand\as{\alpha_s}

\newcommand\LogUV{\,\text{ln}\big[\frac{\mu_0^2}{m_Q^2}\big]}
\newcommand\LogIR{\,\text{ln}\big[\frac{\mu_\Lambda^2}{m_Q^2}\big]}
\newcommand{\mylog}[1]{{\,\text{ln}(#1)}}
\newcommand\logtwo{\,\text{ln}\, 2}

\newcommand{\cc}{{Q\bar{Q}}}
\newcommand{\ME}[2]{\langle\mathcal{O}_{#2}^{[\cc]_\lambda}(#1)\rangle}

\newcommand{\SDC}[4]{\hat{d}^{\,(#1)}_{[\cc(#2)]\to [\cc(#3 #4)]}}
\newcommand{\SDCs}[4]{\hat{d}^{\,(#1)}_{#2\to [\cc(#3 #4)]}}

\newcommand{\ben}{\begin{eqnarray}}
\newcommand{\een}{\end{eqnarray}}

\newcommand{\bef}{\begin{figure}[!htp]}
\newcommand{\eef}{\end{figure}}

\newcommand{\bea}{\begin{eqnarray}}
\newcommand{\eea}{\end{eqnarray}}

\begin{document}

\title{\boldmath Fragmentation functions of polarized heavy quarkonium}

\author[a, b]{Yan-Qing Ma,}
\author[c, d]{Jian-Wei Qiu,}
\author[e]{and Hong Zhang}

\affiliation[a]{Maryland Center for Fundamental Physics,
                   University of Maryland,
                   College Park, Maryland 20742, USA}
\affiliation[b]{Center for High Energy Physics,
                   Peking University,
                   Beijing, 100871, China}
\affiliation[c]{Physics Department,
                   Brookhaven National Laboratory,
                   Upton, New York 11973, USA}
\affiliation[d]{C.N. Yang Institute for Theoretical Physics,
                   Stony Brook University,
                   Stony Brook, New York 11794, USA}
\affiliation[e]{Department of Physics,
                   The Ohio State University,
                   Columbus, Ohio 43210, USA}

\emailAdd{yqma@umd.edu}
\emailAdd{jqiu@bnl.gov}
\emailAdd{zhang.5676@osu.edu}

\date{\today}

\abstract{
Investigating the production of polarized heavy quarkonia in terms of recently proposed
QCD factorization formalism requires the knowledge of a large number of input fragmentation
functions (FFs) from a single parton or a heavy quark-antiquark pair to a polarized heavy
quarkonium.  We study these universal FFs at the input factorization scale $\mu_0\gtrsim 2m_Q$,
with heavy quark mass $m_Q$, in the framework of nonrelativistic QCD (NRQCD) factorization.
We express these FFs in terms of perturbatively calculable coefficients for producing a
heavy quark-antiquark pair in all possible NRQCD states, multiplied by corresponding NRQCD
long-distance matrix elements for the pair to transmute into a polarized heavy quarkonium.
We derive all relevant NRQCD operators for the long-distance matrix elements
based on symmetries, and introduce a self-consistent scheme to define them in arbitrary $d$-dimensions.
We compute, up to the first non-trivial order in $\alpha_s$, the perturbative coefficients
for producing a heavy quark pair in all possible $S$-wave and $P$-wave NRQCD states.
We also discuss the role of the polarized FFs in generating QCD predictions for the polarization
of $J/\psi$ produced at collider energies.
}

\maketitle
\flushbottom

\section{Introduction}

Since the discovery of the $J/\psi$, heavy quarkonia, with their clearly separated multiple momentum scales, have been serving as ideal systems to test our understanding of QCD bound states and their hadronization processes. Unfortunately, a theoretically and phenomenologically satisfying framework is still lacking for heavy quarkonium production. The problem is more acute with the recently discovered $XYZ$-mesons, since the production of these exotic mesons requires good understanding of the production of conventional heavy quarkonia \cite{Brambilla:2010cs,Bodwin:2013nua}.

A phenomenologically successful model for heavy quarkonium production is based on the non-relativistic QCD (NRQCD) factorization \cite{Caswell:1985ui,Bodwin:1994jh}, which factorizes the production cross section into the production of a heavy quark pair, multiplied by the transition for the pair to transmute into the observed quarkonium.  The production of the heavy quark pair is effectively perturbative and is organized in powers of $\alpha_s$ and $v$, the relative velocity of the heavy quark in the pair's rest frame; and the corresponding transition is nonperturbative and is represented by a set of NRQCD long-distance matrix elements (LDMEs). If the factorization is correct to all orders in $\alpha_s$, these LDMEs will be universal, i.e. process independent.
For heavy quarkonium production, although a formal proof of NRQCD factorization to all orders in $\alpha_s$ is still lacking, it has been shown to be valid at the next-to-leading order (NLO) in many processes. With certain modification, it also works at the next-to-next-to-leading order (NNLO) in some specific cases \cite{Nayak:2005rt,Nayak:2005rw,Nayak:2006fm}.  Phenomenologically, NLO NRQCD factorization calculations successfully explain the transverse momentum $p_T$ distribution of many heavy quarkonium states with a few NRQCD LDMEs fitted from the data \cite{Brambilla:2010cs,Bodwin:2013nua}.

Nevertheless, the current NRQCD factorization formalism is far from perfect in describing data on the heavy quarkonium production at high $p_T$.  It is understood now that the fixed-order NRQCD calculation suffers from large high-order corrections, due to the large power enhancement in the forms of $p_T^2/m_Q^2$ and the large $\log(p_T^2/m_Q^2)$-type logarithms from high orders in $\alpha_s$ \cite{Kang:2014tta}, where $m_Q$ is the mass of heavy quark. For example, the NLO NRQCD calculation for the yield of $J/\psi$ is orders of magnitude larger than the leading order (LO) calculation for $\CScSa$ and $\COcPj$ channels at $p_T\gtrsim 15$ GeV \cite{Campbell:2007ws,Gong:2008sn,Ma:2010yw,Butenschoen:2010rq}. Although the most phenomenologically important leading power contribution is claimed to be included in the NLO NRQCD calculation \cite{Ma:2010yw,Ma:2010jj}, the existence of large logarithm may potentially undermine the convergence of $\alpha_s$ expansion at large $p_T\gg m_Q$.

Recently, a new QCD factorization formalism has been proposed to study heavy quarkonium production at large $p_T$ \cite{Kang:2011mg,Kang:2011zza, Fleming:2012wy, Fleming:2013qu, Kang:2014tta, Kang:2014pya}. In this formalism, the cross section is expanded by powers of $m_Q^2/p_T^2$.  It was proved to all orders in $\alpha_s$ that the dominant leading-power (LP) terms, as well as the next-to-leading power (NLP) terms, can be systematically factorized into the perturbatively calculable hard parts for producing a single parton (or a heavy quark pair at NLP), convoluted with corresponding single parton (or heavy quark pair) fragmentation functions (FFs) to the observed heavy quarkonium.  All nonperturbative contributions are included in these FFs, whose scale dependence is determined by a closed set of evolution equations with perturbatively calculable evolution kernels. By solving the evolution equations, large perturbative $\log(p_T^2/m_Q^2)$-type logarithms could be resummed to all orders in $\alpha_s$. Because of the systematic treatment of the powers of $p_T^2/m_Q^2$ and $\log(p_T^2/m_Q^2)$, the QCD factorization is expected to converge faster in $\alpha_s$ expansion than the NRQCD factorization.

With the LO partonic hard parts for hadronic production of heavy quarkonia at high $p_T$ calculated in Ref.~\cite{Kang:2014pya}, and the evolution kernels of FFs available in Refs.~\cite{Fleming:2012wy, Fleming:2013qu,Kang:2014tta}, a set of FFs at some input scale $\mu_0\gtrsim 2m_Q$ is needed and necessary for evaluating the rate of heavy quarkonium production, and for any phenomenological application of the QCD factorization formalism. These input FFs are only unknowns in the QCD factorization framework, and are sensitive to the observed heavy quarkonium states and their polarizations. In principle, these input FFs should be extracted from data, like the case for pion and kaon production at high $p_T$. Nonetheless, this task is extremely hard in practice, especially with the inclusion of NLP contributions. For example, it requires at least four {\it{one-variable}} single parton FFs and six {\it{three-variable}} heavy quark pair FFs for evaluating the production rate of polarization-summed $J/\psi$~\cite{Ma:2013yla,Ma:2014eja}. The number of FFs should double for calculating the cross section of producing polarized $J/\psi$.

However, different from the pion and kaon FFs, the heavy quarkonium FFs have an intrinsic large scale $m_Q\gg \Lambda_\text{QCD}$, thus they are partially perturbative. As early as in 1993, Braaten, Cheung and Yuan proposed to apply the Color-Singlet Model and the NRQCD factorization to further separate the perturbative and nonperturbative contributions to the input FFs~\cite{Braaten:1993mp,Braaten:1993rw}. By choosing the input QCD factorization scale $\mu_0\gtrsim 2m_Q$ and NRQCD factorization scale $\mu_\Lambda \sim m_Q$, neither $\mu_0^2/m_Q^2$ nor $m_Q^2/\mu_\Lambda^2$ is large, and consequently, the NRQCD expansion of the input FFs is expected to have a fast convergence. In Refs.~\cite{Ma:2013yla,Ma:2014eja}, we calculated the polarization-summed input FFs in the framework of the NRQCD factorization.  We derived, up to the NLO in $\alpha_s$, the perturbative hard parts for an energetic single parton (or a heavy quark pair) to fragment into a heavy quark pair in all possible $S$-wave and $P$-wave NRQCD states, while the NRQCD LDMEs cover the  nonperturbative transition rate for the pair to transmute into the observed heavy quarkonium.\footnote{Most single parton input FFs in Refs.~\cite{Ma:2013yla} are not new.  See Refs.~\cite{Ma:2013yla,Ma:2014eja} for the details and references therein.}  These input FFs are needed for evaluating the factorized LP and NLP contributions to the production of polarization-summed heavy quarkonia.

With our calculated input FFs, the first test of the QCD factorization formalism for heavy quarkonium production is shown in Ref.~\cite{Ma:2014svb}. It was found that without the resummation of large logarithms, the simple and fully analytical LO calculation in the QCD factorization approach successfully reproduces the very complicated and numerical NLO NRQCD calculations for the yield of $J/\psi$ at $p_T\gtrsim 15$ GeV, for all phenomenologically important and relevant channels.  It is very important to note that the NLP contribution, although suppressed by a factor of $m_Q^2/p_T^2$ compared to the LP contribution, are crucial for $\COaSz$ and $\CScSa$ channels even when $p_T$ approaches $100$~GeV. A similar calculation with only the LP contribution cannot reproduce these two channels \cite{Bodwin:2014gia}. This finding clearly demonstrates the importance of the NLP contributions to the heavy quarkonium production.

Since the polarization is an important observable for exploring the production mechanism of heavy quarkonium \cite{Butenschoen:2011ks,Chao:2012iv,Gong:2012ug}, it is critically important to study the production of polarized heavy quarkonia in the framework of the QCD factorization formalism, which requires a set of polarized input FFs. In this paper, we extend our work on the polarization-summed input FFs \cite{Ma:2013yla,Ma:2014eja} to the study of the polarized input FFs by using the NRQCD factorization approach.  Different from the production of unpolarized heavy quarkonia, we show that, if the polarization of the produced heavy quarkonium is observed, many more NRQCD channels at the same order in $v$ expansion are needed. Especially, off-diagonal channels are needed for the production of some heavy quarkonia even at the leading power in $v$. We demonstrate that conservation laws, angular momentum addition rules, and velocity scaling rules, as well as the heavy quark symmetry, are very important for constraining the size of contributions from various channels.

Similar to the polarization-summed input FFs \cite{Ma:2013yla,Ma:2014eja}, the intermediate steps in deriving the polarized input FFs require the perturbative treatment of ultraviolet (UV) and infrared (IR) divergences.
Conventional dimensional regularization (CDR) is the most convenient tool to regularize all these divergences. However the application of CDR on the polarized input FFs is not straightforward, since the definitions of polarized NRQCD LDMEs in an arbitrary $d$-dimension are very non-trivial. In principle, one needs to consider the coupling of two $d$-dimensional heavy quark spinors, and L-S coupling if the orbital angular momentum is nonzero, {\it e.g.}, for the $\state{3}{P}{J}{1}$ channels. In this paper, by requiring the symmetries to be preserved when generalizing the polarized NRQCD LDMEs from 4-dimensions to $d$-dimensions, we provide a simple scheme to separate the contributions with different $J$ and $|J_z|$.

The rest of this paper is organized as follows. In Section~\ref{sec:NRQCD}, in terms of the NRQCD factorization formalism, we derive, by using conservation laws and angular momentum addition rules, all nonvanishing NRQCD LDMEs for the production of polarized heavy quarkonia.  We single out the derivation of the nonvanishing NRQCD LDMEs that contribute to polar angular distribution of heavy quarkonium production, which are of the most phenomenological interest. We also discuss how velocity scaling rules help simplify problems in practice. Then in Section~\ref{sec:polarizeddef}, we introduce a scheme to define these polarized NRQCD LDMEs in an arbitrary $d$-dimension. In Section~\ref{sec:InputFFs}, we outline the technical steps needed for calculating the single-parton and heavy quark-antiquark pair ($Q\bar{Q}$ pair) FFs within the NRQCD factorization approach. Since the calculation is very similar to that for the polarization-summed case, which has already been explained in great details in Refs.~\cite{Ma:2013yla,Ma:2014eja}, we do not repeat it in this paper, but providing all mathematical tools specific to the calculation of polarized input FFs in Appendix~\ref{app:LDMEs}. Finally in Section~\ref{sec:summary}, we briefly discuss the impact of our results for the polarized $J/\psi$ production.  All of our results on the polarized input FFs are listed in Appendixes~\ref{app:SinglePolarized} and \ref{app:DoublePolarized}.  In Appendix~\ref{appsubsec:compare}, we compare our full results of single parton FFs with some input FFs available in the literature.

\section{Heavy quarkonium polarization in NRQCD factorization}\label{sec:NRQCD}

In this section, we start from experimental observables to find all allowed NRQCD channels for polarized heavy quarkonium production. In Subsection~\ref{subsec:general}, we drive the most general selection rules for nonvanishing NRQCD channels, by employing only the conservation laws and angular momentum addition rules. These selection rules are valid for any observables related to the heavy quarkonium polarization. Then in Subsection~\ref{subsec:polar}, we focus on the polar angular distribution of decay products of produced heavy quarkonia, and derive more selection rules for this specific observable. Finally in Subsection~\ref{subsec:powercounting}, we adopt the NRQCD power counting rules to find the relative importance of the survived channels for the polar angular distribution.

\subsection{General selection rules}\label{subsec:general}

Experimentally, the polarization of a produced heavy quarkonium is determined by measuring the angular distribution of its decay products in its rest frame. Taking $J/\psi$ as an example, the angular distribution of the decaying $l^+ l^-$ lepton pair can be parameterized as (see Ref.~\cite{Faccioli:2010kd} for a thorough discussion, and  \cite{Braaten:2014ata} for a recent review)
\begin{align}\label{eq:angulardistribution}
\begin{split}
\frac{d\sigma^{J/\psi(\to l^+ l^-)}}{d\Omega} &\propto 1+\lambda_\theta \cos^2\theta+ \lambda_\varphi \sin^2 \theta \cos 2\varphi +\lambda_{\theta\varphi}\sin2\theta \cos\varphi\\
&
+\lambda_\varphi^\perp \sin^2 \theta \sin 2\varphi+\lambda_{\theta\varphi}^\perp \sin2\theta \sin\varphi,
\end{split}
\end{align}
where $\theta$ and $\varphi$ are the polar angle and the azimuthal angle of $l^+$ respect to a chosen $z$-axis of the pair's rest frame, respectively. 

Similarly, one can do the decomposition for any heavy quarkonium state, such as $\chi_{cJ}$ (see \cite{Kniehl:2003pc,Faccioli:2011be,Shao:2012fs}). In the narrow-width approximation for producing a heavy quarkonium, all the coefficients $\lambda$'s in Eq.~(\ref{eq:angulardistribution}) and corresponding coefficients for other heavy quarkonium states can be related to the heavy quarkonium's spin density matrix $\rho^H_{\hspace{-0.1cm}J_z^H,\tilde{J}_z^H}$, where $J_z^H$ ($\tilde{J}_z^H$) is the $z$-component of the spin of the heavy quarkonium $H$ in the amplitude (complex conjugate of the amplitude). Here the quantum interference with $J_z^H\neq \tilde{J}_z^H$ is allowed.
Comparing theoretical predictions of these $\lambda$'s with the data serves as an important test of the theory \cite{Braaten:2008xg,Braaten:2008mz}.

Within the framework of the NRQCD factorization \cite{Bodwin:1994jh}, the heavy quarkonium's spin density matrix can be factorized as
\begin{align}\label{eq:NRQCDFac}
\rho^H_{\hspace{-0.1cm}J_z^H,\tilde{J}_z^H}=\sum_{n,\tilde{n}}\hat{\rho}^{[Q\bar{Q}]}_{n,\tilde{n}} \langle 0 | \mathcal{O}^{H(J_z^H,\tilde{J}_z^H)}_{[Q\bar{Q}(n,\tilde{n})]} |0 \rangle,
\end{align}
where the scale dependence is suppressed. $\hat{\rho}^{[Q\bar{Q}]}_{n,\tilde{n}}$ is the spin density matrix of the polarized intermediate $Q\bar{Q}$ pair with quantum numbers $n=\state{{2S+1}}{L}{J,J_z}{b}$ and $\tilde{n}=\state{{2\tilde{S}+1}}{\tilde{L}}{\tilde{J},\tilde{J}_z}{\tilde{b}}$ in the amplitude and the complex conjugate of the amplitude, respectively, where $S(\tilde{S})$, $L(\tilde{L})$, $J(\tilde{J})$, and $J_z(\tilde{J}_z)$ are spin, orbital angular momentum, total angular momentum and total angular momentum along the $z$-axis  of the pair in the amplitude (complex conjugate of the amplitude), respectively.  Similarly, $b, \tilde{b}=1$ or $8$ represent the color-singlet or color-octet state of the produced pair.  In Eq.~(\ref{eq:NRQCDFac}), the NRQCD LDMEs $\langle 0 | \mathcal{O}^{H({J_z^H,\tilde{J}_z^H})}_{[Q\bar{Q}(n,\tilde{n})]} |0 \rangle$ represent nonperturbative probabilities for the intermediate $[Q\bar{Q}(n,\tilde{n})]$ states to hadronize into the polarized heavy quarkonium $H(J_z^H,\tilde{J}_z^H)$. In principle, for each channel with the definite $n$ and $\tilde{n}$ in Eq.~\eqref{eq:NRQCDFac}, there are still many LDMEs from contributions at different powers of $v^2$. In this section, we do not try to explicitly separate different LDMEs contributing to the same channel in Eq.~\eqref{eq:NRQCDFac}, since we only focus on the selection rules for different channels.

\begin{figure}[htb]
\begin{center}
\includegraphics[width=6cm]{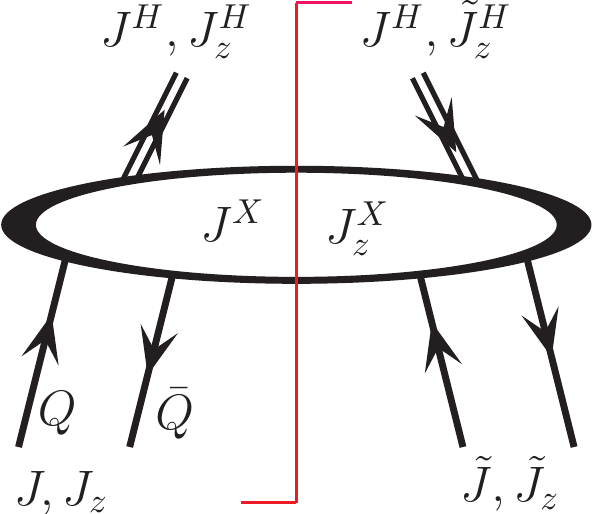}
\caption{Schematic picture of angular momentum coupling in the hadronization from $Q\bar{Q}$-pair to heavy quarkonium $H$.
\label{fig:Jcouple}}
\end{center}
\end{figure}

The summation in Eq.~\eqref{eq:NRQCDFac} runs over all $n$ and $\tilde{n}$, but, the LDMEs $\langle 0 | \mathcal{O}^{H({J_z^H,\tilde{J}_z^H})}_{[Q\bar{Q}(n,\tilde{n})]} |0 \rangle$ vanish for some combinations of $n$ and $\tilde{n}$. A particular channel $[Q\bar{Q}(n,\tilde{n})]$ can contribute to the spin density matrix only when $n$ and $\tilde{n}$ have the same conserved quantum numbers.  For example, $b=\tilde{b}$ and  $(-1)^{L+1}=(-1)^{\tilde{L}+1}$, since the color and parity are always conserved in QED and QCD. With color $b=\tilde{b}$, the state of the heavy quark pair, $[Q\bar{Q}(n,\tilde{n})]$ defined in NRQCD, should be an eigenstate of charge conjugation, which leads to $(-1)^{L+S}=(-1)^{\tilde{L}+\tilde{S}}$.  Combining this with the conservation of color and parity, we obtain the following selection rules:
\begin{align}\label{eq:BSL}
b=\tilde{b},~~~~~~~ S=\tilde{S},~~~~~~~ |L-\tilde{L}|&=0,2,4\cdots \, ,
\end{align}
for the LDMEs $\langle 0 | \mathcal{O}^{H({J_z^H,\tilde{J}_z^H})}_{[Q\bar{Q}(n,\tilde{n})]} |0 \rangle$ in Eq.~(\ref{eq:NRQCDFac}).

Furthermore, angular momentum addition rules can provide more constraints on the choice of $n$ and $\tilde{n}$ for the LDMEs $\langle 0 | \mathcal{O}^{H({J_z^H,\tilde{J}_z^H})}_{[Q\bar{Q}(n,\tilde{n})]} |0 \rangle$.
These nonperturbative NRQCD LDMEs can be represented schematically by the diagram in Fig.~\ref{fig:Jcouple}, which is shown in cut diagram notation, in which the amplitude and its complex conjugate are combined into a forward hadronization diagram for a heavy quark pair to evolve into the heavy quarkonium $H$ and unobserved particles $X$, and the final state is identified by a vertical line.
In Fig.~\ref{fig:Jcouple}, $J^H$ is the total angular momentum of the heavy quarkonium $H$, $J^H_z$ ($\tilde{J}^H_z$) is its $z$-component in the amplitude (complex conjugate of the amplitude), and $J^X$ and $J^X_z$ are the total angular momentum and its $z$-component of the unobserved particles $X$, respectively, which also include the relative orbital angular momentum between $H$ and $X$. By summing over all possible states of $X$, the NRQCD LDMEs, as shown in Fig.~\ref{fig:Jcouple}, can be expressed as
\begin{align}\label{eq:JCouple1}
\begin{split}
\langle 0 | \mathcal{O}^{H(J_z^H,\tilde{J}_z^H)}_{[Q\bar{Q}(n,\tilde{n})]} |0 \rangle
\propto
\hspace{-0.1cm}
\sum_{X}\hspace{-0.1cm}
\sum_{J^X \hspace{-0.1cm},J^X_z}
\hspace{-0.1cm}
A(J^H
\hspace{-0.2cm},
J^X; J, \tilde{J})
\langle \tilde{J}, \tilde{J_z} | J^H
\hspace{-0.1cm}
, \tilde{J}^{H}_z
\hspace{-0.05cm}
;
\hspace{-0.05cm}
J^X
\hspace{-0.1cm}
, J^X_z \rangle
\langle J^H
\hspace{-0.1cm}
, J^H_z
\hspace{-0.05cm}
;
\hspace{-0.05cm}
J^X
\hspace{-0.1cm}
, J^X_z | J, J_z \rangle,
\end{split}
\end{align}
where the coefficient $A(J^H,J^X; J, \tilde{J})$ depends on the dynamics, but, is independent of the $z$-component of any angular momentum due to the rotational invariance of QCD.
Eq.~(\ref{eq:JCouple1}) leads to the following constraints,
\begin{align}\label{eq:JJzRelation}
|J-\tilde{J}| &\leq 2J^H,
{\text{ and }}
J_z-\tilde{J}_z=J_z^H-\tilde{J}^H_z\, .
\end{align}
Eqs.~(\ref{eq:BSL}) and (\ref{eq:JJzRelation}) summarize all the constraints for the NRQCD LDMEs in Eq.~(\ref{eq:NRQCDFac}) for the production of a polarized $H$ that decays into two bodies, with the measured full angular distribution of the decaying products, like the $J/\psi$-production in Eq.~(\ref{eq:angulardistribution}).

The constraints in Eq.~(\ref{eq:JJzRelation}) can be modified if the angular distribution of the decay products is integrated.  As we will show in the next subsection, the polar angle $\theta$-distribution (after the integration of the azimuthal angle of the decaying products) depends only on the diagonal entries of the heavy quarkonium spin density matrix $\rho^H_{\hspace{-0.1cm}J_z^H,{J}_z^H}$.  Thus, only the diagonal LDMEs $\langle 0 | \mathcal{O}^{H(J_z^H,J_z^H)}_{[Q\bar{Q}(n,\tilde{n})]} |0 \rangle$, with $\tilde{J}_z^H={J}^H_z$, are needed for the $\theta$-distribution of heavy quarkonium production, and therefore, the constraints in Eq.~(\ref{eq:JJzRelation}) can be simplified to
\begin{align}\label{eq:JJzRelation1}
|J-\tilde{J}| &\leq 2J^H,
{\text{ and }}
J_z=\tilde{J}_z,
\hspace{1cm}
(\text{for the polar $\theta$-distribution}).
\end{align}
For the yield of a polarization-summed heavy quarkonium $H$, we integrate over both the polar $\theta$ and azimuthal $\varphi$ distribution of the decay products, and consequently, we only need $\langle 0 | \mathcal{O}^{H}_{[Q\bar{Q}(n,\tilde{n})]} |0 \rangle\equiv \sum_{J_z^H} \langle 0 | \mathcal{O}^{H(J_z^H,J_z^H)}_{[Q\bar{Q}(n,\tilde{n})]} |0 \rangle$. By choosing $\tilde{J}^H_z=\, J_z^H$ and summing over $J^H_z$ in Eq.~(\ref{eq:JCouple1}), the orthonormality condition of angular momentum requires
\begin{align}\label{eq:JJzRelation2}
J=\tilde{J},
{\text{ and }}
J_z=\tilde{J}_z,
\hspace{1cm}
(\text{for producing a polarization-summed $H$}).
\end{align}
That is, for the yield of a polarization-summed $H$, only the fully diagonal entries in both $J$ and $J_z$ of the heavy quarkonium spin density matrix are needed.

\subsection{Nonvanishing channels for the polar angular distribution}\label{subsec:polar}

In this paper, we focus on the polar angular distribution, which is of the most phenomenological interest. For $J/\psi$ production, the polar angular distribution means the $\lambda_\theta$ term in Eq.~(\ref{eq:angulardistribution}).  By integrating the azimuthal angle $\varphi$ from 0 to $2\pi$ on both sides of Eq.~(\ref{eq:angulardistribution}), we obtain
\begin{align}\label{eq:angulardistribution1}
\begin{split}
\frac{d\sigma^{J/\psi(\to l^+ l^-)}}{d\cos \theta} &\propto 1+\lambda_\theta \cos^2\theta.
\end{split}
\end{align}
In the narrow-width approximation, the polar angular distribution for producing the $J/\psi$ that decays into a lepton $l^+ l^-$ pair can be expressed as
\begin{align}\label{eq:NWAres}
\frac{d\sigma^{J/\psi(\to l^+ l^-)}}{d\cos \theta}=\sum_{J_z^{J/\psi}, \tilde{J}_z^{J/\psi}}
\rho^{J/\psi}_{J_z^{J/\psi}, \tilde{J}_z^{J/\psi}} \times
\epsilon_{{i}}({J}_z^{J/\psi}) \mathcal{M}_\text{decay}^{i,j} \epsilon^*_j(\tilde{J}_z^{J/\psi})
\end{align}
where $\mathcal{M}_\text{decay}$ is the decay matrix element of $J/\psi$, and $\epsilon_i(J_z^{J/\psi})$ and $\epsilon^*_{\tilde{i}}(\tilde{J}_z^{J/\psi})$ are the polarization vectors of the $J/\psi$ in the decay amplitude and its complex conjugate, respectively.  Due to the parity conservation and the rotational invariance about the $z$-axis after the integration of azimuthal angle, $\varphi$, the decay matrix for the polar angular distribution must have the structure $\mathcal{M}_\text{decay}^{i,j} = c_1~\delta^{i,j} + c_2~\delta^{i,z}\delta^{j,z}$, where $c_1$ and $c_2$ are the coefficients determined by the dynamics. That is, only the diagonal entries, i.e. the terms with $J_z^{J/\psi}=\tilde{J}_z^{J/\psi}$ of the $J/\psi$ spin density matrix, $\rho^H_{J_z^H,\tilde{J}_z^H}$ in Eq.~(\ref{eq:NRQCDFac}), can contribute to the polar angular distribution in Eq.~\eqref{eq:NWAres} \cite{Braaten:2014ata}.

Furthermore, because of the parity and time-reversal invariance of QED, we have $\epsilon_{{i}}({J}_z^{J/\psi}) \mathcal{M}_\text{decay}^{i,j} \epsilon^*_j({J}_z^{J/\psi})=\epsilon_{{i}}(-{J}_z^{J/\psi}) \mathcal{M}_\text{decay}^{i,j} \epsilon^*_j(-{J}_z^{J/\psi})$. Consequently, the polar angular distribution depends only on the combination of $\rho^{J/\psi}_{\hspace{-0.1cm}J_z^{J/\psi},J_z^{J/\psi}}+\rho^{J/\psi}_{\hspace{-0.1cm}-J_z^{J/\psi},-J_z^{J/\psi}}$.  Similar argument also applies to the production of other heavy quarkonium states. That is, the polar angular distribution of the decaying products from any produced heavy quarkonium $H$ depends only on the combination of the spin density matrices: $\rho^H_{\hspace{-0.1cm}J_z^H,J_z^H}+\rho^H_{\hspace{-0.1cm}-J_z^H,-J_z^H}$.

Therefore, we can use the selection rules in Eqs.~(\ref{eq:BSL}) and (\ref{eq:JJzRelation1}) to find out all relevant $Q\bar Q$-pair channels $(n,\tilde{n})$ that contribute to the polar angular distribution. In addition to all diagonal channels with $n=\tilde{n}$, there are also many offdiagonal channels. For example, if $n$ is $S$-wave, all $S$-$D$ mixing channels are allowed; and if $n$ is $P$-wave, there are $P$-$P$ mixing channels as well as $P$-$F$ mixing channels:
\begin{align}\label{eq:offdiagonal}
\begin{split}
&\left(\state{1}{S}{0}{b},\state{1}{D}{2,0}{b}\right), \hspace{0.5cm} \left(\state{3}{S}{1,J_z}{b},\state{3}{D}{\tilde{J},J_z}{b}\right), \hspace{0.5cm}\left(\state{3}{P}{J,J_z}{b},\state{3}{P}{\tilde{J},J_z}{b}\right), \\
&\left(\state{1}{P}{1,J_z}{b},\state{1}{F}{3,J_z}{b}\right) \hspace{0.5cm} \left(\state{3}{P}{J,J_z}{b},\state{3}{F}{\tilde{J},J_z}{b}\right),
\end{split}
\end{align}
where $b=1,8$ labels the color of the $Q\bar{Q}$ pair.

Since the polar angular distribution is proportional to the sum of two spin density matrices with the opposite signed $z$-component of quarkonium angular momentum, we introduce the reduced spin density matrices as
\begin{align}\label{eq:spindensity2}
\rho^H_{|J_z^H|} \equiv
\frac{1}{2}\left(
\rho^H_{\hspace{-0.1cm}J_z^H,J_z^H}+\rho^H_{\hspace{-0.1cm}-J_z^H,-J_z^H}
\right)
\equiv
\rho^H_{\lambda}
\end{align}
with $\lambda=L, T, TT, \cdots$, corresponding to $|J_z^H|=0, 1, 2, \cdots$, respectively.
From the NRQCD factorization in Eq.~\eqref{eq:NRQCDFac}, and using
$\langle 0 | \mathcal{O}^{H(J_z^H,J_z^H)}_{[Q\bar{Q}(n,\tilde{n})]} |0 \rangle=\langle 0 | \mathcal{O}^{H(-J_z^H,-J_z^H)}_{[Q\bar{Q}(-\tilde{n}, -n)]} |0 \rangle$, where $-n\equiv \state{{2S+1}}{L}{J,-J_z}{b}$ and similarly for $-\tilde{n}$, which is an immediate consequence of the time-reversal invariance of QCD,
we find that the reduced spin density matrices in Eq.~(\ref{eq:spindensity2}) can be factorized as,
\begin{align}\label{eq:NRQCDFac2}
\begin{split}
\rho^H_{\lambda}=&\sum_{n,\tilde{n}}\frac{1}{2}\left( \hat{\rho}^{[Q\bar{Q}]}_{n,\tilde{n}}+\hat{\rho}^{[Q\bar{Q}]}_{-\tilde{n},-n}\right)
\frac{1}{2}\left( \langle 0 | \mathcal{O}^{H_\lambda}_{[Q\bar{Q}(n,\tilde{n})]} |0 \rangle + \langle 0 | \mathcal{O}^{H_\lambda}_{[Q\bar{Q}(-\tilde{n},-n)]} |0 \rangle \right),
\end{split}
\end{align}
where the reduced NRQCD operators are defined as
\begin{align}
\mathcal{O}^{H_\lambda} \equiv
\overline{\sum_{|J_z^H|}}\mathcal{O}^{H(J_z^H,J_z^H)}
=
\frac{1}{2} \left(
\mathcal{O}^{H(J_z^H,J_z^H)}+\mathcal{O}^{H(-J_z^H,-J_z^H)}
\right)\, ,
\end{align}
where the subscript $\lambda$ is defined right below Eq.~(\ref{eq:spindensity2}).

\subsection{NRQCD power counting}\label{subsec:powercounting}

The general discussions in last subsection were based only on the symmetry properties of QED and QCD. In principle, one needs all allowed $Q\bar{Q}$ channels for any quarkonium production. In practice, however, different channels are not equally important. The relative importance of different channels in power of $v$ is governed by the velocity scaling rule of NRQCD \cite{Bodwin:1994jh}.

\begin{table}
\begin{center}
\caption{Essential channels for various polarized heavy quarkonium production, with relative power-counting of each channel explicitly.  $\lambda$ represents the polarization of the heavy quarkonium, and could be $L$, $T$ or $TT$ depending on the specific state. $Q$ can be either charm or bottom.
\label{tab:pc4fermionPolarized}}
\vspace{0.3cm}
\begin{tabular}{c |  l}
\hline \hline
Quarkonium &  Essential channels \\
\hline
$\eta_Q(nS)$ & $(\state{1}{S}{{0}}{1})[v^0]$, $(\state{3}{S}{{1,L}}{8})[v^3]$,  $(\state{3}{S}{{1,T}}{8})[v^3]$ \\
\hline
$\begin{array}{l} \psi(nS), \Upsilon(nS) \end{array}$ &
$ \begin{array}{l}
(\state{3}{S}{{1,\lambda}}{1})[v^0], (\COaSz)[v^3], (\state{3}{S}{{1,\lambda}}{8})[v^4],
(\state{3,\lambda}{P}{}{8})[v^4]
\end{array} $\\
\hline
$h_Q(nP)$ & $(\state{1}{P}{{1,\lambda}}{1})[v^0]$, $(\COaSz)[v^0]$, $(\state{3}{S}{{1,L}}{8})[v^3]$, $(\state{3}{S}{{1,T}}{8})[v^3]$ \\
\hline
$\chi_{QJ}(nP)$ &
$\begin{array}{l} (\state{3}{P}{{J,\lambda}}{1})[v^0], (\state{3}{S}{{1,L}}{8})[v^0], (\state{3}{S}{{1,T}}{8})[v^0] \end{array}$\\
\hline \hline
\end{tabular}
\end{center}
\end{table}

To derive the scaling rule, one first expands the wave function of a physical heavy quarkonium $H$ into different Fock states in NRQCD effective field theory. For a conventional heavy quarkonium, its dominant Fock state is a $Q\bar{Q}$-pair Fock state with a definite quantum number, which is the same as the one usually used to denote the produced quarkonium, while all other Fock states are suppressed by powers of $v$. The relative suppression can be estimated by color multipole expansion: the suppression is at $O(v)$ if two Fock states are related by a color electric dipole transition (E1); and the suppression is at $O(v^{3/2})$ if the two Fock states are related by a color magnetic dipole transition (M1) \cite{Bodwin:1994jh}. Therefore, the expansion for the wave function of a heavy quarkonium $H$ with quantum number $\statenc{2S+1}{L}{J}$ can be expressed as,
\begin{align}
\begin{split}
|H(\statenc{2S+1}{L}{J})\rangle = & O(1)|Q\bar{Q}(\state{2S+1}{L}{J}{1})\rangle +  O(v)|Q\bar{Q}(\state{2S+1}{(L\pm1)}{J'}{8})g\rangle\\
& + O(v^{3/2})|Q\bar{Q}(\state{3-2S}{L}{J'}{8})g\rangle +  O(v^2)|Q\bar{Q}(\state{2S+1}{L'}{J'}{1,8})gg\rangle+\cdots.
\end{split}
\end{align}
In addition to the suppression caused by the Fock states expansion, there is another suppression factor at $O(v^{L+\tilde{L}})$ for the $Q\bar Q$-pair channel $(n,\tilde{n})$ because of the derivative operation associated with the orbital angular momentum of the operators. Based on these power counting rules, one can easily work out the relative importance in $v$ of different channels for the production of a given heavy quarkonium.  We list the relative power counting of phenomenologically important channels for $S$-wave and $P$-wave quarkonium production in Table \ref{tab:pc4fermionPolarized}.
Three general conclusions are in order:
\begin{itemize}
\item First, a channel mixing between different waves is always suppressed by at least $O(v^2)$. For example, the $\left(\state{1}{S}{0}{b},\state{1}{D}{2,0}{b}\right)$ channel in Eq.~\eqref{eq:offdiagonal} is always suppressed by $O(v^2)$ comparing with $\left(\state{1}{S}{0}{b},\state{1}{S}{0}{b}\right)$ channel for any quarkonium state.

\item Second, a channel mixing between different total angular momentum but the same orbital angular momentum can be expressed in terms of a ``diagonal'' channel up to $O(v^2)$ correction. For example, the mixing channels $\left(\state{3}{P}{J,J_z}{8},\state{3}{P}{\tilde{J},J_z}{8}\right)$ in Eq.~\eqref{eq:offdiagonal} contributes to the production of $H=J/\psi$, $\psi(nS)$ or $\Upsilon(ns)$. If we assume the heavy quark spin symmetry, the spin of the intermediate $Q\bar{Q}$-pair in the amplitude, $S_z$, and that in the complex-conjugate amplitude, $\tilde{S}_z$, should equal to the spin of the hadron $J_z^{H}$. Furthermore, the NRQCD LDMEs corresponding to the mixing channel and the channel with $S_z=\tilde{S}_z=J_z^H$ are equal to each other due to heavy quark spin symmetry.  As a result, we can define a ``diagonal'' channel $\left(\state{3,S_z}{P}{}{8},\state{3,S_z}{P}{}{8}\right)$ to study heavy quarkonium production, in which the spin of the $Q\bar{Q}$-pair is $S_z$, and the orbital angular momentum is summed over. As heavy quark spin symmetry holds up to $O(v^2)$ correction \cite{Bodwin:1994jh}, this approximate treatment is valid at the same order of the precision.

\item Third, for any heavy quarkonium state, there is at least one $S$-wave channel which contributes at leading power in $v$; for any $P$-wave and higher-wave heavy quarkonium state, there is at least one $P$-wave channel which contributes at leading power in $v$; and so on. Thus lower-wave channels are usually more important.
\end{itemize}
With this consideration, we will calculate only $S$-wave and $P$-wave channels in this paper, including the ``diagonal'' channels $\left(\state{3,S_z}{P}{}{b},\state{3,S_z}{P}{}{b}\right)$ reduced from $P$-$P$ mixing channels in Eq.~\eqref{eq:offdiagonal}, and leave a complete treatment of offdiagonal channels and higher-wave channels for future study.
Since it involves only diagonal channels in our treatment (up to $O(v^2)$ corrections), the factorization formula in Eq.~\eqref{eq:NRQCDFac2} can be further simplified to
\begin{align}\label{eq:NRQCDFac3}
\begin{split}
\rho^H_{\lambda}\approx&
\sum_{n}\frac{1}{2}\left( \hat{\rho}^{[Q\bar{Q}]}_{n,{n}}+\hat{\rho}^{[Q\bar{Q}]}_{-{n},-n}\right)
\frac{1}{2}\left( \langle 0 | \mathcal{O}^{H_\lambda}_{[Q\bar{Q}(n,{n})]} |0 \rangle + \langle 0 | \mathcal{O}^{H_\lambda}_{[Q\bar{Q}(-{n},-n)]} |0 \rangle \right)
\\
\equiv&
\sum_{n_{\lambda'}} \hat{\rho}^{[Q\bar{Q}]}_{n_{\lambda'}} \,
\langle 0 | \mathcal{O}^{H_\lambda}_{[Q\bar{Q}(n_{\lambda'})]} |0 \rangle,
\end{split}
\end{align}
where, similar to the parameter $\lambda$ defined in Eq.~\eqref{eq:spindensity2} for the $z$-components of heavy quarkonium angular momentum $J^H$, we introduced a parameter, $\lambda'=L, T, TT, \cdots$ representing the $z$-component of the $Q\bar{Q}$-pair's angular momentum, $|J_z|=0, 1, 2, \cdots$, respectively. For example,
\begin{align}
\begin{split}
\hat{\rho}^{[Q\bar{Q}]}_{n_{T}}=&
\sum_{n(|J_z|=1)}\frac{1}{2}\left( \hat{\rho}^{[Q\bar{Q}]}_{n,{n}}+\hat{\rho}^{[Q\bar{Q}]}_{-{n},-n}\right)
\equiv
\overline{\sum_{|J_z|=1}}\hat{\rho}^{[Q\bar{Q}]}_{(n,n)}\hspace{0.5cm}, \\
\langle 0 | \mathcal{O}^{H_\lambda}_{[Q\bar{Q}(n_{T})]} |0 \rangle=&
\sum_{n(|J_z|=1)}
\frac{1}{2}\left( \langle 0 | \mathcal{O}^{H_\lambda}_{[Q\bar{Q}(n,{n})]} |0 \rangle + \langle 0 | \mathcal{O}^{H_\lambda}_{[Q\bar{Q}(-{n},-n)]} |0 \rangle \right) \\
\equiv&
\overline{\sum_{|J_z|=1}} \langle 0 | \mathcal{O}^{H_\lambda}_{[Q\bar{Q}(n,n)]} |0 \rangle.
\end{split}
\end{align}
To relate our results in this paper to polarization-summed results, we sum over the polarization $\lambda$ in Eq.~\eqref{eq:NRQCDFac3}, and obtain
\begin{align}\label{eq:NRQCDFacAve}
\begin{split}
\sum_\lambda \rho^H_{\lambda}\approx&\sum_{n_{\lambda'}} \hat{\rho}^{[Q\bar{Q}]}_{n_{\lambda'}} \sum_\lambda\langle 0 | \mathcal{O}^{H_\lambda}_{[Q\bar{Q}(n_{\lambda'})]} |0 \rangle={\sum_{n_{\lambda'}}} \hat{\rho}^{[Q\bar{Q}]}_{n_{\lambda'}}\frac{1}{2J+1}\langle 0 | \mathcal{O}^{H}_{[Q\bar{Q}(\bar{n})]} |0 \rangle,
\end{split}
\end{align}
where $\langle 0 | \mathcal{O}^{H}_{[Q\bar{Q}(\bar{n})]} |0 \rangle=\sum_{{J_z}}\langle 0 | \mathcal{O}^{H}_{[Q\bar{Q}(n,n)]} |0 \rangle$ is the usual polarization-summed NRQCD LDME. To obtain the second equal sign in Eq.~\eqref{eq:NRQCDFacAve}, we have used the factor that the summation $\sum_\lambda\langle 0 | \mathcal{O}^{H_\lambda}_{[Q\bar{Q}(n_{\lambda'})]} |0 \rangle$ is independent of $\lambda'$ because of rotation invariance.

Our calculations in this paper should be sufficient for the current phenomenological study of $S$-wave and $P$-wave heavy quarkonium production.  In the next section, we provide a simple scheme to define all these operators in $d$-dimensions.

\section{Definition of polarized NRQCD LDMEs in $d$ dimensions}\label{sec:polarizeddef}

In Refs.~\cite{Ma:2013yla,Ma:2014eja}, we calculated both the single parton and heavy quark pair FFs for an polarization-summed heavy quarkonium. We used CDR to regularize all the UV and IR divergences. By doing this, we implicitly generalized the polarization-summed NRQCD LDMEs to arbitrary $d$-dimensions. With $J_z$  of the heavy quarkonium summed over, this generalization is simple since there is no special direction in the heavy quarkonium rest frame. In this polarization-summed case, the possible generalization can be achieved by extending the $SO(3)$ symmetry to the $SO(d-1)$ symmetry~\cite{Ma:2013yla,Ma:2014eja}.  That is, we demand all $Q\bar{Q}$-pair channels to be covariant under the action of $SO(d-1)$ rotation group, similar to the 4-dimensions case.

The situation is more complicated with the quarkonium's polarization observed, where a specific ${z}$-axis direction needs to be specified. For example, in the hadron {\it helicity} frame, the ${z}$-axis is chosen to be along the moving direction of the heavy quarkonium in the Laboratory frame. To separate contributions with the same $J$ but different $|J_z|$, we need to know more details of the $SO(d-1)$ rotation group. The situation could be more severe with angular momentum couplings, such as the L-S coupling, which is exactly what we have in the NRQCD factorization.

In this section, we provide a simple scheme to separate the contributions with different $J$ and $|J_z|$, which is required for the calculation of the polar angle distribution of the decay products, like the $\lambda_\theta$ in Eq.~(\ref{eq:angulardistribution1}). In our scheme, we only require: (1) the wave function of the heavy quark pair preserves all the symmetries about the ${z}$-axis when it is generalized to $d$-dimensions, and (2) the $d$-dimensional polarization-summed NRQCD LDMEs in Ref.~\cite{Ma:2013yla,Ma:2014eja} are recovered after adding up corresponding polarized ones. Consequently, the following rules are valid for $S$-wave and $P$-wave channels,
\begin{itemize}
\item
Case 1: for $\statenc{3}{S}{1}$ and $\statenc{1}{P}{1}$, the wave functions of the heavy quark pair with $|J_z|=1$ (or $J_z=0$) in its rest frame are antisymmetric (or symmetric) when flipping the direction of all axes except $\hat{z}$ axe. Note that, it is hard to define the rotation operation in $d$ dimensions, but the flipping operation described here is always well-defined\footnote{We effectively assume that in both $4$-dimensions and $d$-dimensions, the operation of rotating $\pi$ rad about the ${z}$-axis is well defined, and is equivalent to flipping the direction of all axes except $z$-axis.}.
\item
Case 2: for $\statenc{3}{P}{J}$, the wave function with $J=0$ is a scalar, i.e. it has a $SO(d-1)$ symmetry. Wave functions with $J=2$ (or $J=1$) are symmetric (or anti-symmetric) tensors in their orbital and spin indices, and are constructed to be traceless.
\item
Case 3: for $\statenc{3}{P}{2}$, the wave functions with $|J_z|=1$ can be separated from those with $|J_z|=0,2$ by requiring that they are antisymmetric under flipping the direction of all axes except ${z}$, similar to the Case 1 above. The wave functions with $J_z=0$ and $|J_z|=2$ can be further separated by requiring the wave function with $J_z=0$ to have a $SO(d-2)$ symmetry in the space perpendicular to $z$-axis.
\item
Case 4: for $\statenc{3}{P}{1}$, the wave functions with $|J_z|=1$ can be singled out by requiring that they are antisymmetric under flipping the direction of all axes except ${z}$, similar to the Case 1 above. We define the symmetric part under this operation as the wave function with $J_z=0$.\footnote{Different from the $4$-dimension case, wave functions with $|J_z|=0,1,2$ for $\statenc{3}{P}{1}$ channel are allowed in $d$-dimensions. Wave functions with $|J_z|=1$ are transversely polarized at $d=4$. Wave functions with $|J_z|=2$ have different parity from wave functions with $|J_z|=1$, and vanish at $d=4$.  It is then natural to consider $|J_z|=2$ wave functions as longitudinally polarized in $d$-dimensions.}
\end{itemize}
Notice that in $d$-dimensions, there is no unique way to group states into categories with different $J$ and $|J_z|$. Different grouping methods are equally good as long as they are consistent and give the correct decomposition at $d=4$. They could serve as different schemes in the dimensional regularization.

As an example, we consider the polarized LDMEs for $J/\psi$ production below, which is of great phenomenological interest, while giving definitions of all $S$-wave and $P$-wave NRQCD LDMEs in Appendix~\ref{app:LDMEs}. The most important $[Q\bar{Q}(n)]$ channels in Eq.~(\ref{eq:NRQCDFac3}) for $J/\psi$ production are $\CScSa$, $\COcSa$, $\COaSz$ and $\COcPj$ (see Table.~\ref{tab:pc4fermionPolarized}). For $\CScSa$ channel, the wave function for polarization-summed $Q\bar{Q}$-pair in its rest frame can be chosen as a $(d-1)$ dimensional vector\footnote{The coupling of two spinors in $d$-dimensions also have high-spin representations. On the one hand, their contribution to spin-0 and spin-1 representation is at least suppressed by $v^3$ because of the heavy quark spin symmetry \cite{Petrelli:1997ge}. On the other hand, one could always ignore them by using a different dimensional regularization scheme, where spinors are in $4$-dimensions.

Here we only consider the components at leading powers in $v$, relativistic corrections can be constructed by inserting covariant derivatives, as in Ref.~\cite{Bodwin:1994jh}.}
\begin{align}\label{eq:3s1summed}
|\Psi_{Q\bar{Q}[\statenc{3}{S}{1}]}  \rangle = c^j \psi^\dagger \sigma^{j} \chi |0 \rangle
\end{align}
where $\psi$ ($\chi$) is the annihilation (the creation) operator for a heavy quark (a heavy antiquark) in the center of mass frame of the pair, $\{ \psi^\dagger \sigma^{j} \chi |0 \rangle \} $ with $j=1\dots (d-1)$ are the basis states/vectors of the Hilbert space for $|\Psi_{Q\bar{Q}[\statenc{3}{S}{1}]} \rangle$, and $c^j$ are the corresponding coordinates. In this Hilbert space with $d=4$, the $3\times 3$ matrix representation of the flipping operation described in the Case 1 above is ${diag}(-1,-1,1)$, which can be easily generalized to $(d-1)\times(d-1)$ matrix representation:
\begin{align}\label{eq:FlipRep}
\text{D}^\text{flip}_{i,j}={diag}(-1,\cdots, -1,1),
\end{align}
where ellipsis represents $(d-4)$ copies of $-1$.
With Eq.~(\ref{eq:FlipRep}), we separate the ``longitudinal polarization'' space spanned by the basis state $\{ \psi^\dagger \sigma^{z} \chi |0 \rangle \} $ from the ``transverse polarization'' space spanned by the basis states $\{ \psi^\dagger \sigma^{j_\bot} \chi |0 \rangle \} $, with $j_\bot$ includes all values of $j$ defined in Eq.~(\ref{eq:3s1summed}), except $j=z$.
We could then construct the definitions of polarized NRQCD LDMEs for $[Q\bar{Q}(\CScSa)]$ in $d$-dimensions as
\begin{subequations}\label{eq:CScSaTL}
\begin{align}
\begin{split}\label{eq:CScSaT}
\mathcal{O}^{H_\lambda}(\CScSaT)
&=\frac{1}{d-2}\frac{1}{2N_c}\chi^\dagger \sigma^{j_\perp} \psi (a_{H_\lambda}^\dagger a_{H_\lambda}) \psi^\dagger \sigma^{j_\perp} \chi,
\end{split}\\
\begin{split}\label{eq:CScSaL}
\mathcal{O}^{H_\lambda}(\CScSaL)
&= \frac{1}{2N_c}\chi^\dagger \sigma^z \psi (a_{H_\lambda}^\dagger a_{H_\lambda}) \psi^\dagger \sigma^z \chi,
\end{split}
\end{align}
\end{subequations}
where the summation of $j_\perp$ is suppressed, $a_{H_\lambda}$ is the annihilation operator of heavy quarkonium $H$ with polarization $\lambda$, and $\frac{1}{(d-2)(2N_c)}$ is a normalization factor. The corresponding NRQCD LDMEs for $\COcSa$ channel can be obtained by removing $1/(2N_c)$ in Eq.~(\ref{eq:CScSaTL}) and inserting one SU(N) color matrices $T^a$ to the operator of each heavy quark pair.

For $J/\psi$ production, the analysis of the $\COcPj$ channel is greatly simplified if we assume the heavy quark spin symmetry, as explained at the end of Subsection~\ref{subsec:powercounting}. Since the spin of the produced $J/\psi$ must be the same as the spin of the intermediate $Q\bar{Q}$-pair in the leading power of $v$, the orbital angular momentum of the heavy quark pair can be effectively summed over. In this way, we can apply the same flipping operation as in Eq.~(\ref{eq:FlipRep}), but only to the spin index, and we obtain
\begin{subequations}\label{eq:jpsi3pj}
\begin{align}
\begin{split}
\mathcal{O}^{H_\lambda}(\COcPT)
&=\frac{1}{(d-1)(d-2)}\chi^\dagger(-\frac{i}{2}\overleftrightarrow{D}^k \sigma^{j_\perp})\,T^a\psi (a_{H_\lambda}^\dagger a_{H_\lambda}) \psi^\dagger (-\frac{i}{2}\overleftrightarrow{D}^k \sigma^{j_\perp}) \,T^a \chi,
\end{split}\\
\begin{split}
\mathcal{O}^{H_\lambda}(\COcPL)
&=\frac{1}{(d-1)}\chi^\dagger(-\frac{i}{2}\overleftrightarrow{D}^j \sigma^{z})\,T^a\psi (a_{H_\lambda}^\dagger a_{H_\lambda}) \psi^\dagger (-\frac{i}{2}\overleftrightarrow{D}^j \sigma^{z}) \,T^a \chi,
\end{split}
\end{align}
\end{subequations}
where
\begin{align}
\begin{split}
\psi^\dagger\overleftrightarrow{\boldsymbol{D}}\chi
&\equiv
\psi^\dagger(\boldsymbol{D}\chi)-(\boldsymbol{D}\psi)^\dagger\chi.
\end{split}
\end{align}
A complete treatment of the $\state{3}{P}{J}{8}$ channel without summing over the orbital angular momentum needs $Q\bar{Q}$ channels offdiagonal in $(n, \tilde{n})$, as analyzed in Section~\ref{sec:NRQCD}. For simplicity, we do not consider individual offdiagonal LDMEs as their overall contributions can be included in the LDMEs in Eq.~(\ref{eq:jpsi3pj}).

\section{NRQCD factorization on input FFs}\label{sec:InputFFs}

The NRQCD factorization formula derived in Eq.~\eqref{eq:NRQCDFac3} can also be applied to FFs at the initial scale $\mu_0$. To be specific, the input single parton FFs and the $Q\bar{Q}$-pair FFs to a polarized heavy quarkonium can be written in the following factorized form\footnote{In this paper we work in the hadron {\it helicity} frame, in which the mixed fragmentation with a single parton in the amplitude and a heavy quark pair in the complex conjugate of the amplitude does not exist, following the same argument in Section IIB of Ref.~\cite{Kang:2014tta}}
\begin{subequations}\label{eq:FF}
\begin{align}
\begin{split}\label{eq:singleFF}
D_{f\to H_{\lambda'}}(z;m_Q,\mu_0)
=
\sum_{n_{\lambda}} \hat{d}_{f\to[\cc(n_{\lambda})]}(z;m_Q,\mu_0,\mu_\Lambda)
\langle \mathcal{O}_{[\cc(n_{\lambda})]}^{H_{\lambda'}}(\mu_\Lambda)\rangle,
\end{split}\\
\begin{split}\label{eq:QQbarFF}
\mathcal{D}_{[Q\bar{Q}(\kappa)]\to H_{\lambda'}}(z,\zeta_1,\zeta_2;m_Q,\mu_0)
\hspace{0cm}
=
\sum_{n_{\lambda}}
\hspace{-0.cm}
 \hat{d}_{[Q\bar{Q}](\kappa)\to[\cc(n_{\lambda})]}(z,\zeta_1,\zeta_2;m_Q,\mu_0,\mu_\Lambda)
\langle \mathcal{O}_{[\cc(n_{\lambda})]}^{H_{\lambda'}}(\mu_\Lambda)\rangle,
\end{split}
\end{align}
\end{subequations}
where $\mu_0\gtrsim 2m_Q$ is the QCD factorization scale, $\mu_\Lambda\sim m_Q$ is the NRQCD factorization scale, and $\lambda$ ($\lambda'$) denotes the polarization of the intermediate NRQCD $Q\bar{Q}$ pair (observed heavy quarkonium $H$).  In Eq.~(\ref{eq:singleFF}), $f=Q, \bar{Q}, q,\bar{q},g$ is the flavor of the fragmenting parton, and the variable $z$ is the light-cone momentum fraction of the parton taken by the quarkonium $H$.  In Eq.~(\ref{eq:QQbarFF}), $\kappa= v^{[1,8]}, a^{[1,8]}$ or $t^{[1,8]}$ represents the vector, axial-vector or tensor spin states of the fragmenting heavy quark pair, respectively, where the superscript labels the singlet (1) or octet (8) color state of the pair.  The variables $\zeta_1$ and $\zeta_2$ in Eq.~(\ref{eq:QQbarFF}), defined in Refs.~\cite{Ma:2013yla,Ma:2014eja}, represent the relative light-cone momentum fractions of the heavy quark pair in the amplitude and its complex conjugate, respectively.

The short-distance coefficients (SDCs), $\hat{d}$'s in Eq.~(\ref{eq:FF}), are insensitive to any particular quarkonium state $H_{\lambda'}$, which is an immediate consequence of the factorization.  We can derive these SDCs by replacing the final heavy quarkonium state by an asymptotic heavy quark pair $[Q\bar{Q}(n'_{\lambda'})]$,
\begin{subequations}\label{eq:FF1}
\begin{align}
\begin{split}\label{eq:singleFF1}
\hspace{-0.3cm}
D_{f\to [Q\bar{Q}(n'_{\lambda'})]}(z;m_Q)
&=
\sum_{n_{\lambda}} \hat{d}_{f\to[\cc(n_{\lambda})]}(z;m_Q)
\langle \mathcal{O}_{[\cc(n_{\lambda})]}^{[Q\bar{Q}(n'_{\lambda'})]}\rangle,
\end{split}\\
\begin{split}\label{eq:QQbarFF1}
\hspace{-0.3cm}
\mathcal{D}_{[Q\bar{Q}](\kappa)\to [Q\bar{Q}(n'_{\lambda'})]}(z,\zeta_1,\zeta_2;m_Q)
&=
\hspace{-0.cm}
\sum_{n_{\lambda}}
\hspace{-0.cm}
 \hat{d}_{[Q\bar{Q}](\kappa)\to[\cc(n_{\lambda})]}(z,\zeta_1,\zeta_2;m_Q)
\langle \mathcal{O}_{[\cc(n_{\lambda})]}^{[Q\bar{Q}(n'_{\lambda'})]}\rangle,
\end{split}
\end{align}
\end{subequations}
where the dependence on the factorization scales is suppressed. In Eq.~(\ref{eq:FF1}), the LHS can be calculated with perturbative QCD, while the LDMEs on the RHS can be calculated with perturbative NRQCD. By matching the LHS and the RHS, the SDCs can be obtained order by order in $\alpha_s$.

In this paper, we follow the convention used in Refs.~\cite{Ma:2013yla,Ma:2014eja}, and expand the SDCs as
\begin{subequations}
\begin{align}
\begin{split}\label{eq:NRQCDFacSingle}
D_{f \to H_{\lambda'}}(z;m_Q,\mu_0)
=&
\sum_{n_{\lambda}}
\pi \as \Big\{\hat{d}^{\,(1)}_{f \to [\cc(n_{\lambda})]}(z;m_Q,\mu_0,\mu_\Lambda)\\
&+ \left(\frac{\as}{\pi}\right)\,\hat{d}^{\,(2)}_{f \to [\cc(n_{\lambda})]}(z;m_Q,\mu_0,\mu_\Lambda)+O(\alpha_s^2)\Big\}
\times
\frac{{\langle \mathcal{O}_{[\cc(n_{\lambda})]}^{H_{\lambda'}}(\mu_\Lambda)\rangle}}{m_Q^{2L+3}},
\end{split}
\end{align}
\begin{align}
\begin{split}\label{eq:NRQCDFacDouble}
{\mathcal D}_{[\cc(\kappa)]\to H_{\lambda'}}(z,\zeta_1,\zeta_2,\mu_0;m_Q)
&=
\sum_{n_{\lambda}}
\Big\{\hat{d}^{\,(0)}_{[\cc(\kappa)]\to [\cc(n_{\lambda})]}(z,\zeta_1,\zeta_2,\mu_0;m_Q,\mu_\Lambda)
\\
&\hspace{-3cm}+
\left(\frac{\as}{\pi}\right)
\hat{d}^{\,(1)}_{[\cc(\kappa)]\to [\cc(n_{\lambda})]}(z,\zeta_1,\zeta_2,\mu_0;m_Q,\mu_\Lambda)
+O(\alpha_s^2)\Big\}
\times
\frac{\langle \mathcal{O}_{[\cc(n_{\lambda})]}^{H_{\lambda'}}(\mu_\Lambda)\rangle}{m_Q^{2L+1}}.
\end{split}
\end{align}
\end{subequations}

Now with the $d$-dimensional polarized NRQCD LDMEs and the projection operators derived from these LDMEs, we can apply CDR to regularize all the divergences in the calculation of the LHS of Eq.~(\ref{eq:FF1}), with a particular polarized $[Q\bar{Q}(n'_{\lambda'})]$  state. In the NLO perturbative calculation of the LHS, there exists (1) UV divergences, which are removed by QCD renormalization and the renormalization of composite operators; (2) rapidity divergence, which are canceled when adding up all Feynman diagrams; (3) Coulomb divergence and IR divergence, which are canceled between the LHS and the NRQCD LDMEs on the RHS, if NRQCD factorization is valid. The cancellation of the Coulomb divergence and IR divergence needs NLO calculation of polarized NRQCD LDMEs, which are also given in Appendix~\ref{app:LDMEs}.

These polarized SDCs and the polarization-summed SDCs are not independent. With the definitions of the polarized NRQCD LDMEs in Appendix~\ref{app:LDMEs}, one can obtain the relation
\begin{subequations}\label{eq:PoUnPoRelation}
\begin{align}
\hat{d}^{\,(1 \ or\  2)}_{f \to [\cc(n)]}
=&\frac{1}{N^\text{NR}_{n}}\sum_\lambda
\hat{d}^{\,(1\ or\ 2)}_{f\to [\cc(n_{\lambda})]},
\\
\hat{d}^{\,(0 \ or\  1)}_{[\cc(\kappa)]\to [\cc(n)]}
=&\frac{1}{N^\text{NR}_{n}}\sum_\lambda
\hat{d}^{\,(0\ or\ 1)}_{[\cc(\kappa)]\to [\cc(n_{\lambda})]},
\end{align}
\end{subequations}
where $\hat{d}^{\,(1\ or \ 2)}_{f \to [\cc(n)]}$ and $\hat{d}^{\,(0\ or \ 1)}_{[\cc(\kappa)]\to [\cc(n)]}$ are the polarization-summed SDCs, and $N^\text{NR}_{n}$ are number of polarization states for the channel $n$, both of which can be found in Refs.~\cite{Ma:2013yla,Ma:2014eja}. The summation of $\lambda$ runs over all polarization states of $n$. In Appendix \ref{app:SinglePolarized} and \ref{app:DoublePolarized}, we only list the $\hat{d}^{\,(1\ or \ 2)}_{f\to [\cc(n_{\lambda})]}$ and $\hat{d}^{\,(0 \ or\  1)}_{[\cc(\kappa)]\to [\cc(n_{\lambda})]}$ for one polarization state for $\statenc{3}{S}{1}$, $\statenc{1}{P}{1}$, and $\statenc{3}{P}{1}$ channels, and two polarization states for $\statenc{3}{P}{2}$ channel. One can derive the rest by using the relations in Eq.~(\ref{eq:PoUnPoRelation}).

\section{Discussion and Summary}\label{sec:summary}

In this paper, we derived all nonvanishing channels of producing a heavy-quark pair $(n,\tilde{n})$ for polarized heavy quarkonium production in NRQCD factorization. We introduced a scheme to generalize the polarized NRQCD LDMEs to arbitrary $d$-dimensions. With these $d$-dimensional NRQCD LDMEs, we apply CDR to calculate the single-parton and heavy-quark pair FFs, up to the NLO in $\alpha_s$, for the production of all polarized $S$-wave and $P$-wave heavy quarkonium at the input scale $\mu_0\gtrsim 2m_Q$. We find that all perturbative divergences are canceled at this order, and all derived SDCs are finite as expected. With our results, as well as the evolution kernels and the hard parts calculated in Refs.~\cite{Kang:2014tta,Kang:2014pya}, the QCD factorization formalism can be used to generate predictions for the polarization of produced heavy quarkonia in the hadron {\it helicity} frame, which have been measured at the Tevatron and the LHC.

In addition to the surprisingly large NLP contribution to the yield of $J/\psi$ \cite{Ma:2014svb}, our results in this paper further demonstrate the potential importance of the NLP to the $J/\psi$ polarization.  In Table~\ref{tab:PolarizedFFsJpsi}, we show the LO contributions from different channels to the $J/\psi$ polarization. The single-quark fragmentations are not important since they are suppressed at large $z$ \cite{Ma:2013yla}. The contribution with a fragmenting $\cc$ pair in a tensor state is suppressed by the hard part \cite{Kang:2014pya}. These suppressed channels are not showed in Table~\ref{tab:PolarizedFFsJpsi}. For completeness, we also list the $\COaSz$ channel, which contributes to unpolarized $J/\psi$ production.

\begin{table}[!htb]
\centering
\caption{Contributions of LO FFs to the $J/\psi$ polarization. The labels ``T'', ``L'', and ``Un'' represent transversely polarized, longitudinally polarized, and unpolarized $J/\psi$, respectively.
\label{tab:PolarizedFFsJpsi}}
\vspace{0.3cm}
\begin{tabular}{c | c | c | c | c}
\hline \hline
 & $\CScSa$ & $\COcSa$ & $\COcPj$ & $\COaSz$ \\
\hline
g &  & T & & \\
\hline
$\vone$ & L & & & \\
\hline
$\veight$ & & L & L & \\
\hline
$\aone$ & & & & \\
\hline
$\aeight$ & & & T & Un\\
\hline \hline
\end{tabular}
\end{table}

Contrary to the contribution at the LP, most production channels at the NLP contribute to longitudinally polarized $J/\psi$. Recall that the NLP contributes to the production rate, both directly by the NLP term in the factorized cross section, and indirectly through the mixed kernels in the evolution equations of the single-parton FFs (see Ref.~\cite{Kang:2014tta} for more details). Especially, the indirect contribution can increase the longitudinal component of produced $J/\psi$ from gluon fragmentation when the gluon FF is evolved from the input scale $\mu_0\gtrsim 2m_Q$ to the hard scale $\mu\sim p_T$. Although the details need more studies, we believe our results in this paper will be very helpful for understanding the polarization of heavy quarkonium production.

\section*{Acknowledgments}

We thank E.~Braaten and G.~Sterman for helpful discussions and G.T.~Bodwin for useful communication regarding comparisons between our work and the results in Ref. \cite{Bodwin:2014bia}. HZ would like to thank the hospitality of the Peking University.
This work was supported in part by the U. S. Department of Energy under contract Nos. DE-AC02-98CH10886 and DE-SC0011726, Office of Nuclear Physics under Award Number DE-FG02-93ER-40762,
and the National Science Foundation under grant Nos.~PHY-0354776, PHY-0354822 and PHY-0653342.

\appendix
\section{Polarized NRQCD LDMEs}\label{app:LDMEs}

This appendix is organized as follows. In Subsection~\ref{subsec:essential}, we list the definitions of all essential $d$-dimensional polarized NRQCD LDMEs with the $S$-wave and $P$-wave heavy-quark pair. In Subsection~\ref{subsec:polarizedproj}, we present the projection operators for the calculation of SDCs. In Subsection~\ref{subsec:polarizedLDMENLO}, we expand the polarized NRQCD LDMEs to NLO in $\alpha_s$, which are necessary for the full cancellation of IR divergences in the NLO calculation of SDCs.

\subsection{NRQCD LDMEs}\label{subsec:essential}

With the method expained in Section~\ref{sec:polarizeddef}, we give our definitions of normalized NRQCD four-fermion operators for polarized heavy quarkonium production in an arbitrary dimension $d$.
\begin{subequations}\label{eq:Polarized4fermion}
\begin{align}
\begin{split}
\mathcal{O}^{H_\lambda}(\COcSaT)
&=\frac{1}{(d-2)}\chi^\dagger \sigma^{j_\perp} \,T^a\psi (a_{H_\lambda}^\dagger a_{H_\lambda}) \psi^\dagger \sigma^{j_\perp} \, T^a\chi,
\end{split}\\
\begin{split}
\mathcal{O}^{H_\lambda}(\COcSaL)
&= \chi^\dagger \sigma^z\,T^a\psi (a_{H_\lambda}^\dagger a_{H_\lambda}) \psi^\dagger \sigma^z\, T^a\chi,
\end{split}\\
\begin{split}
\mathcal{O}^{H_\lambda}(\COaPaT)
&=\frac{1}{(d-2)} \chi^\dagger(-\frac{i}{2}\overleftrightarrow{D}^{j_\perp})\,T^a\psi (a_{H_\lambda}^\dagger a_{H_\lambda}) \psi^\dagger (-\frac{i}{2} \overleftrightarrow{D}^{j_\perp})\,T^a \chi,
\end{split}\\
\begin{split}
\mathcal{O}^{H_\lambda}(\COaPaL)
&= \chi^\dagger(-\frac{i}{2}\overleftrightarrow{D}^z)\,T^a\psi (a_{H_\lambda}^\dagger a_{H_\lambda}) \psi^\dagger (-\frac{i}{2} \overleftrightarrow{D}^z)\, T^a \chi,
\end{split}\\
\begin{split}
\mathcal{O}^{H_\lambda}(\COcPaT)
&=\frac{1}{2(d-2)} \chi^\dagger(-\frac{i}{2}\overleftrightarrow{D}^{\,\,[\,j_\perp}\, \sigma^{z]})\,T^a\psi (a_{H_\lambda}^\dagger a_{H_\lambda}) \psi^\dagger (-\frac{i}{2}\overleftrightarrow{D}^{\,\,[\,j_\perp}\, \sigma^{z]}) \,T^a \chi,
\end{split}\\
\begin{split}
\mathcal{O}^{H_\lambda}(\COcPaL)
&= \frac{1}{2(d-2)(d-3)}\,\chi^\dagger(-\frac{i}{2}\overleftrightarrow{D}^{\,\,[\,j_\perp}\, \sigma^{k_\perp]})\,T^a\psi (a_{H_\lambda}^\dagger a_{H_\lambda}) \psi^\dagger (-\frac{i}{2}\overleftrightarrow{D}^{\,\,[\,j_\perp}\, \sigma^{k_\perp]}) \,T^a \chi,
\end{split}\\
\begin{split}
\mathcal{O}^{H_\lambda}(\COcPbTb)
&= \frac{2}{(d-1)(d-2)-2}\,\chi^\dagger(-\frac{i}{2}(\frac{1}{2}\overleftrightarrow{D}^{\,\,\{\,j_\perp}\, \sigma^{k_\perp\}}-\frac{\delta^{j_\perp k_\perp}}{d-2}\overleftrightarrow{\boldsymbol{D}}_T\cdot \boldsymbol{\sigma}_T))\,T^a\psi
\\
&\hspace{1cm}
(a_{H_\lambda}^\dagger a_{H_\lambda})
\psi^\dagger (-\frac{i}{2}(\frac{1}{2}\overleftrightarrow{D}^{\,\,\{\,j_\perp}\, \sigma^{k_\perp\}}-\frac{\delta^{j_\perp k_\perp}}{d-2}\overleftrightarrow{\boldsymbol{D}}_T\cdot \boldsymbol{\sigma}_T)) \, T^a\chi,
\end{split}\\
\begin{split}
\mathcal{O}^{H_\lambda}(\COcPbTa)
&=\frac{1}{2(d-2)}\,\chi^\dagger(-\frac{i}{2}\overleftrightarrow{D}^{\,\,\{\,j_\perp}\, \sigma^{z\}})\,T^a\psi
(a_{H_\lambda}^\dagger a_{H_\lambda})
\psi^\dagger (-\frac{i}{2}\overleftrightarrow{D}^{\,\,\{\,j_\perp}\, \sigma^{z\}}) \, T^a\chi,
\end{split}\\
\begin{split}
\mathcal{O}^{H_\lambda}(\COcPbL)
&=\frac{d-2}{d-1}\,\chi^\dagger(-\frac{i}{2}(\overleftrightarrow{D}^z\, \sigma^{z}-\frac{1}{d-2}\overleftrightarrow{\boldsymbol{D}}_T\cdot \boldsymbol{\sigma}_T))\,T^a\psi
\\
&\hspace{1cm}
(a_{H_\lambda}^\dagger a_{H_\lambda})
\psi^\dagger (-\frac{i}{2}(\overleftrightarrow{D}^z\, \sigma^{z}-\frac{1}{d-2}\overleftrightarrow{\boldsymbol{D}}_T\cdot \boldsymbol{\sigma}_T)) \, T^a\chi,
\end{split}\\
\begin{split}
\mathcal{O}^{H_\lambda}(\COcPz)
&=\frac{1}{d-1}\,\chi^\dagger(-\frac{i}{2}\overleftrightarrow{\boldsymbol{D}}\cdot \boldsymbol{\sigma})\,T^a\psi
(a_{H_\lambda}^\dagger a_{H_\lambda})
\psi^\dagger (-\frac{i}{2}\overleftrightarrow{\boldsymbol{D}}\cdot \boldsymbol{\sigma}) \, T^a\chi,
\end{split}\\
\begin{split}
\mathcal{O}^{H_\lambda}(\COaSz)
&= \chi^\dagger\,T^a
\psi
(a_{H_\lambda}^\dagger a_{H_\lambda})
\psi^\dagger \,T^a \chi,
\end{split}
\end{align}
\end{subequations}
where the summations of $j_\perp$ and $k_\perp$ run over all directions perpendicular to $\hat{z}$-axis, and
\begin{subequations}
\begin{align}
\begin{split}
\psi^\dagger\overleftrightarrow{\boldsymbol{D}}\chi
&\equiv
\psi^\dagger(\boldsymbol{D}\chi)-(\boldsymbol{D}\psi)^\dagger\chi,
\end{split}\\
\begin{split}
\overleftrightarrow{\boldsymbol{D}}_T\cdot \boldsymbol{\sigma}_T
&\equiv
\overleftrightarrow{\boldsymbol{D}}\cdot \boldsymbol{\sigma}-\overleftrightarrow{{D}}^z {\sigma}^z,
\end{split}\\
\begin{split}
\overleftrightarrow{D}^{\,\,[\,j_\perp}\, \sigma^{k_\perp]}
&\equiv
\overleftrightarrow{D}^{j_\perp}\, \sigma^{k_\perp}-\overleftrightarrow{D}^{k_\perp}\, \sigma^{j_\perp},
\end{split}\\
\begin{split}
\overleftrightarrow{D}^{\,\,\{\,j_\perp}\, \sigma^{k_\perp\}}
&\equiv
\overleftrightarrow{D}^{j_\perp}\, \sigma^{k_\perp}+\overleftrightarrow{D}^{k_\perp}\, \sigma^{j_\perp}.
\end{split}
\end{align}
\end{subequations}
Subscripts $TT$, $T$, and $L$ represent the non-relativistic $\cc$ pair with $|J_z|=2$, $|J_z|=1$ and $J_z=0$, respectively. Polarization of the heavy quarkonium $H$ is labelled by $\lambda$, which can be $TT$, $T$ and $L$, depending on the specific $J_z^H$ state of the $H$. For completeness, we also list the operators for states $\statenc{3}{P}{0}$ and $\statenc{1}{S}{0}$, which are unpolarized.

As mentioned at the end of Section~\ref{sec:polarizeddef}, the $\COcPj$ channel can be greatly simplified by using the heavy quark spin symmetry, so that we only need two NRQCD LDMEs in Eq.~(\ref{eq:jpsi3pj}):
\begin{subequations}\label{eq:jpsi3pjapp}
\begin{align}
\begin{split}
\mathcal{O}^{H_\lambda}(\COcPT)
&=\frac{1}{(d-1)(d-2)}\chi^\dagger(-\frac{i}{2}\overleftrightarrow{D}^k \sigma^{j_\perp})\,T^a\psi (a_{H_\lambda}^\dagger a_{H_\lambda}) \psi^\dagger (-\frac{i}{2}\overleftrightarrow{D}^k \sigma^{j_\perp}) \,T^a \chi,
\end{split}\\
\begin{split}
\mathcal{O}^{H_\lambda}(\COcPL)
&=\frac{1}{(d-1)}\chi^\dagger(-\frac{i}{2}\overleftrightarrow{D}^j \sigma^{z})\,T^a\psi (a_{H_\lambda}^\dagger a_{H_\lambda}) \psi^\dagger (-\frac{i}{2}\overleftrightarrow{D}^j \sigma^{z}) \,T^a \chi,
\end{split}
\end{align}
\end{subequations}
The color-singlet operators can be obtained from their color-octet counterparts in Eqs.~(\ref{eq:Polarized4fermion}) and (\ref{eq:jpsi3pjapp}) by removing the two $T^a$'s and multiplying the factor $1/(2N_c)$.

The LDMEs defined in Eqs.~\eqref{eq:Polarized4fermion} and \eqref{eq:jpsi3pjapp} have been normalized by the number of spin states,
\begin{subequations}\label{eq:NormPolarized}
\begin{align}
N^{\text{NR}}_{\statenc{3}{S}{1,T}}&=N^{\text{NR}}_{\statenc{1}{P}{1,T}}=N^{\text{NR}}_{\statenc{3}{P}{1,T}}=N^{\text{NR}}_{\statenc{3}{P}{2,T}}=d-2,\\
N^{\text{NR}}_{\statenc{3}{S}{1,L}}&=N^{\text{NR}}_{\statenc{1}{P}{1,L}}=N^{\text{NR}}_{\statenc{3}{P}{2,L}}=1,\\
N^{\text{NR}}_{\statenc{3}{P}{1,L}}&=\frac{1}{2}(d-2)(d-3),\\
N^{\text{NR}}_{\statenc{3}{P}{2,TT}}&=
\frac{1}{2}(d-1)(d-2)-1,\\
N^{\text{NR}}_{\statenc{3,T}{P}{}}&=
(d-1)(d-2),\\
N^{\text{NR}}_{\statenc{3,L}{P}{}}&=
d-1.
\end{align}
\end{subequations}
By adding the number of spin states with the same $J$ but different $|J_z|$, we can retrieve the normalization factor for unpolarized heavy-quark-pair in Ref.~\cite{Ma:2013yla,Ma:2014eja}. With the definitions above, it is straightforward to check that adding up all operators with the same $J$ but different $|J_z|$ weighted by the number of spin states, we reproduce the conventional definitions of unpolarized NRQCD LDMEs at arbitrary $d$ dimension,
\begin{align}
\begin{split}
\langle \mathcal{O}_{[\cc(n)]}^{H_{\lambda}}(\mu_\Lambda)\rangle=
\sum_{n_{\lambda'}} N^{\text{NR}}_{n_{\lambda'}} \langle \mathcal{O}_{[\cc(n_{\lambda'})]}^{H_{\lambda}}(\mu_\Lambda)\rangle,
\end{split}
\end{align}
which correspond to definitions in Ref.~\cite{Bodwin:1994jh} by setting $d=4$. \footnote{\label{footnote:diffdef}Note that our definitions of color singlet NRQCD LDMEs differ from the definitions in Ref.~\cite{Bodwin:1994jh} by a factor of $1/(2N_c)$, while the definitions of color octet NRQCD LDMEs are the same.}

\subsection{Projection Operators}\label{subsec:polarizedproj}

With the definitions of $d$-dimensional polarized NRQCD LDMEs, we derive the projection operators in the same way as in the polarization-summed case,
\begin{subequations}
\begin{align}
P^{\text{NR}}_{\statenc{3}{S}{1,T}}&=P^{\text{NR}}_{\statenc{1}{P}{1,T}}=\mathbb{P}_\perp^{\beta \beta'}(p),\\
P^{\text{NR}}_{\statenc{3}{S}{1,L}}&=P^{\text{NR}}_{\statenc{1}{P}{1,L}}=\mathbb{P}_\parallel^{\beta \beta'}(p),\\
P^{\text{NR}}_{\statenc{3}{P}{1,T}}&=\frac{1}{2}\left(
\mathbb{P}_\perp^{\alpha \alpha'}(p) \mathbb{P}_\parallel^{\beta \beta'}(p)
+\mathbb{P}_\perp^{\beta \beta'}(p) \mathbb{P}_\parallel^{\alpha \alpha'}(p)
-\mathbb{P}_\perp^{\alpha \beta'}(p)\mathbb{P}_\parallel^{\beta \alpha'}(p)\right.\nonumber\\
&\hspace{1cm}
\left.
-\mathbb{P}_\perp^{\beta \alpha'}(p) \mathbb{P}_\parallel^{\alpha \beta'}(p)
\right),\\
P^{\text{NR}}_{\statenc{3}{P}{1,L}}&=\frac{1}{2}\left(
\mathbb{P}_\perp^{\alpha \alpha'}(p) \mathbb{P}_\perp^{\beta \beta'}(p)
-\mathbb{P}_\perp^{\alpha \beta'}(p)\mathbb{P}_\perp^{\beta \alpha'}(p)
\right),\\
P^{\text{NR}}_{\statenc{3}{P}{2,TT}}&=
\frac{1}{2}\left(\mathbb{P}_\perp^{\alpha \alpha'}(p) \mathbb{P}_\perp^{\beta \beta'}(p)+\mathbb{P}_\perp^{\alpha \beta'}(p)\mathbb{P}_\perp^{ \alpha' \beta}(p)\right)
-\frac{1}{d-2}\mathbb{P}_\perp^{\alpha \beta}(p)\mathbb{P}_\perp^{\alpha' \beta'}(p),\\
P^{\text{NR}}_{\statenc{3}{P}{2,T}}&=\frac{1}{2}\left(
\mathbb{P}_\perp^{\alpha \alpha'}(p) \mathbb{P}_\parallel^{\beta \beta'}(p)
+\mathbb{P}_\perp^{\beta \beta'}(p)\mathbb{P}_\parallel^{\alpha \alpha'}(p)
+\mathbb{P}_\perp^{\alpha \beta'}(p)\mathbb{P}_\parallel^{\beta \alpha'}(p)\right.\nonumber\\
&\hspace{1cm}
\left.
+\mathbb{P}_\perp^{\beta \alpha'}(p)\mathbb{P}_\parallel^{\alpha \beta'}(p)
\right),\\
P^{\text{NR}}_{\statenc{3}{P}{2,L}}&=\frac{d-2}{d-1}
\large(\mathbb{P}_\parallel^{\alpha \beta}(p)-\frac{1}{d-2}\mathbb{P}_\perp^{\alpha\beta}(p)\large)
\large(\mathbb{P}_\parallel^{\alpha' \beta'}(p)-\frac{1}{d-2}\mathbb{P}_\perp^{\alpha' \beta'}(p)\large),\\
P^{\text{NR}}_{\statenc{3}{P}{0}}&=\frac{1}{d-1}
\mathbb{P}^{\alpha \beta}(p)\mathbb{P}^{\alpha' \beta'}(p),\\
P^{\text{NR}}_{\statenc{1}{S}{0}}&=1,\\
P^{\text{NR}}_{\statenc{3,T}{P}{}}&=\mathbb{P}^{\alpha \alpha'}(p)\mathbb{P}_\perp^{\beta \beta'}(p),\\
P^{\text{NR}}_{\statenc{3,L}{P}{}}&=\mathbb{P}^{\alpha \alpha'}(p)\mathbb{P}_\parallel^{\beta \beta'}(p),
\end{align}
\end{subequations}
where the superscript ``NR'' refers to NRQCD, and
\begin{subequations}
\begin{align}
\begin{split}
\mathbb{P}_\perp^{\alpha \alpha'}(p)&=
-g^{\alpha \alpha'}+\frac{p^\alpha \hat{n}^{\alpha'}+p^{\alpha'} \hat{n}^\alpha}{p\cdot \hat{n}}-\frac{p^2}{(p\cdot \hat{n})^2}\hat{n}^\alpha \hat{n}^{\alpha'},
\end{split}\\
\begin{split}
\mathbb{P}_\parallel^{\alpha \alpha'}(p)&=
\frac{p^\alpha p^{\alpha'}}{p^2}-\frac{p^\alpha \hat{n}^{\alpha'}+p^{\alpha'} \hat{n}^\alpha}{p\cdot \hat{n}}+\frac{p^2}{(p\cdot \hat{n})^2}\hat{n}^\alpha \hat{n}^{\alpha'},
\end{split}\\
\begin{split}
\mathbb{P}^{\alpha \alpha'}(p)&=
\mathbb{P}_\parallel^{\alpha \alpha'}(p)+\mathbb{P}_\perp^{\alpha \alpha'}(p)
=-g^{\alpha\alpha'}+\frac{p^\alpha p^{\alpha'}}{p^2}.
\end{split}
\end{align}
\end{subequations}
$\alpha$ and $\beta$ ($\alpha'$ and $\beta'$) are the indices for the orbital angular momentum and spin of the heavy quark pair in the amplitude (complex conjugate of the amplitude), respectively. By adding up all projection operators with different $|J_z|$ but the same $J$, we can get the same unpolarized projection operators in Refs.~\cite{Ma:2013yla,Ma:2014eja}. By setting $d=4$ for the $P^{\text{NR}}_{\statenc{3}{P}{2,\lambda}}$ with $\lambda=L, T, TT$, we retrieve the results in Ref.~\cite{kniehl:2000nn}.

\subsection{Expand the polarized LDMEs to NLO in perturbative NRQCD}\label{subsec:polarizedLDMENLO}

To complete the cancelation of all IR divergences in the NLO calculation of SDCs, we need to expand the polarized NRQCD LDMEs to NLO in powers of $\alpha_s$, the same as what was done in the polarization-summed case. We calculate the NLO NRQCD corrections for four-fermion operators using similar method as described in Ref.~\cite{Fan:2009cj}. For our purpose, we only give the NLO correction of 4-fermion S-wave operators:
\begin{subequations}\label{eq:MEExpand}
\begin{align}
\begin{split}
\ME{\CSaSz}{}
&=
\ME{\CSaSz}{}^{\text{LO}}\\
&\hspace{0.5cm}
-C_\epsilon \frac{1}{2N_c}\Big\{(d-2)\ME{\COaPaT}{}^\text{LO}+\ME{\COaPaL}{}^\text{LO} \Big\},
\end{split}\\
\begin{split}
\ME{\COaSz}{}
&=
\ME{\COaSz}{}^{\text{LO}}\\
&\hspace{0.5cm}
-C_\epsilon \Big\{ C_F \left((d-2)\ME{\CSaPaT}{}^\text{LO}+\ME{\CSaPaL}{}^\text{LO} \right)\\
&\hspace{1.5cm}
+B_F \left((d-2)\ME{\COaPaT}{}^\text{LO}+\ME{\COaPaL}{}^\text{LO} \right)\Big\},
\end{split}\\
\begin{split}
\ME{\CScSaT}{}
&=
\ME{\CScSaT}{}^{\text{LO}}
-C_\epsilon \frac{1}{2N_c}\,W_T^{[8]},
\end{split}\\
\begin{split}
\ME{\CScSaL}{}
&=
\ME{\CScSaL}{}^{\text{LO}}
-C_\epsilon \frac{1}{2N_c}\,W_L^{[8]},
\end{split}\\
\begin{split}
\ME{\COcSaT}{}
&=
\ME{\COcSaT}{}^{\text{LO}}
-C_\epsilon \Big\{ C_F\,W_T^{[1]}+B_F \, W_T^{[8]}\Big\},
\end{split}\\
\begin{split}
\ME{\COcSaL}{}
&=
\ME{\COcSaL}{}^{\text{LO}}
-C_\epsilon \Big\{ C_F \, W_L^{[1]}+B_F \, W_L^{[8]}\Big\},
\end{split}
\end{align}
\end{subequations}
where $C_F=\frac{N_c^2-1}{2N_c}$, $B_F=\frac{N_c^2-4}{4N_c}$ and $C_\epsilon
=\frac{4\alpha_s}{3\pi m_Q^2}\frac{1}{\epsilon_{IR}}(4\pi e^{-\gamma_E})^\epsilon\left(\frac{\mu_r}{\mu_\Lambda}\right)^\epsilon$. In Eq.~\eqref{eq:MEExpand}, $W_L^{[1,8]}$ and $W_T^{[1,8]}$ are defined as
\begin{subequations}\label{eq:CSCO}
\begin{align}
\begin{split}
W_T^{[b]}
=&(d-1)\ME{\CBcPT}{}^\text{LO}+\cdots\\
=&
\frac{1}{d-1}\ME{\CBcPz}{}^\text{LO}+\frac{1}{2}\ME{\CBcPaT}{}^\text{LO}\\
&+\frac{d-3}{2}\ME{\CBcPaL}{}^\text{LO}+\frac{(d-1)(d-2)-2}{2(d-2)}\ME{\CBcPbTb}{}\\
&+\frac{1}{2}\ME{\CBcPbTa}{}^\text{LO}+\frac{1}{(d-1)(d-2)}\ME{\CBcPbL}{}^\text{LO}+\cdots,
\end{split}\\
\nonumber\\
\begin{split}
W_L^{[b]}
=&(d-1)\ME{\CBcPL}{}^\text{LO}+\cdots\\
=&
\frac{1}{d-1}\ME{\CBcPz}{}^\text{LO}+\frac{d-2}{2}\ME{\CBcPaT}{}^\text{LO}\\
&
+\frac{d-2}{2}\ME{\CBcPbTa}{}^\text{LO}+\frac{d-2}{d-1}\ME{\CBcPbL}{}^\text{LO}
+\cdots,
\end{split}
\end{align}
\end{subequations}
where $\langle{\cal O}^{[Q\bar{Q}]_\lambda}(^{3,T}P^{[b]})\rangle^{\text{LO}}$ and 
$\langle{\cal O}^{[Q\bar{Q}]_\lambda}(^{3,L}P^{[b]})\rangle^{\text{LO}}$ are the leading order terms 
of LDMEs of NRQCD operators in Eqs~ (\ref{eq:jpsi3pjapp}), respectively, and ellipsis denotes terms that are irrelevant for our calculation.

To derive the expressions in Eqs.~\eqref{eq:MEExpand}, we replace $d$ by $(4-2\epsilon)$ only for the dimension of loop momentum in the calculation while keeping $d$'s in other places untouched, and define a renormalization scheme for LDMEs to subtract all terms that are proportional to $C_\epsilon^{UV}$, which is the same as $C_\epsilon$ by replacing $\epsilon_{IR}$ by $\epsilon_{UV}$. Our renormalization scheme is similar to the conventional $\overline{\text{MS}}$ scheme, but not exactly the same. The advantage of our renormalization scheme is that, if LDMEs are defined differently by a factor of $1+O(\epsilon)$, the $d$ dependence in Eqs.~\eqref{eq:MEExpand} will be modified correspondingly, but the calculated finite short-distance coefficients given in the next two Appendixes are unaltered .

\section{Single-Parton Fragmentation Functions to a Polarized Heavy Quarkonium}
\label{app:SinglePolarized}

In this appendix we list the SDCs for all single-parton fragmentation functions  to $S$-wave and $P$-wave polarized $\cc$-pair up to order $O(\alpha_s^2)$ defined in Eq.~(\ref{eq:NRQCDFacSingle}). The polarized FFs and unpolarized FFs are related by Eq.~(\ref{eq:PoUnPoRelation}). Therefore for outgoing $\cc$ with $n+1$ polarizations, we only give $n$ polarized FFs below. The other one can be derived by using Eq.~(\ref{eq:PoUnPoRelation}) and the unpolarized FFs calculated in Ref.~\cite{Ma:2013yla}. All channels that vanish for all polarizations are not listed.

\subsection{Gluon FFs}
{\bf Leading order}
\begin{align}
\SDCs{1}{g}{\COcSaL}{}&=0.
\end{align}
{\bf Next-to-leading order}
\begin{align}
\begin{split}\label{eq:gto3P11L}
\SDCs{2}{g}{\CScPaL}{}&=
\frac{1}{3N_c}
\Big\{
\delta(1-z) \Big[ -\LogIR + 2 \logtwo +\frac{1}{2} \Big]\\
&\hspace{1cm}
+z(2z^2-z+1)\frac{1}{(1-z)_+}
\Big\},
\end{split}\\
\begin{split}\label{eq:gto3P21L}
\SDCs{2}{g}{\CScPbL}{}&=
\frac{1}{9 N_c z^4}
\Big\{
\delta(1-z) \Big[ -\LogIR + 2 \logtwo +\frac{1}{2} \Big]\\
&\hspace{1cm}
+2z^4\frac{1}{(1-z)_+}
-216(z-2)(z-1)^2\mylog{1-z}
\\
&\hspace{1cm}
-z(2z^5+5z^4+38z^3-468z^2+864z-432)
\Big\},
\end{split}\\
\begin{split}\label{eq:gto3P21T}
\SDCs{2}{g}{\CScPbTa}{}&=
\frac{1}{3 N_c z^4}
\Big\{
\delta(1-z) \Big[ -\LogIR + 2 \logtwo  \Big]
+2z^4\frac{1}{(1-z)_+}\\
&\hspace{1cm}
-48(z^4-5z^3+10z^2-10z+4)\mylog{1-z}
\\
&\hspace{1cm}
-2z(z^5+4z^4-55z^3+152z^2-192z+96)
\Big\},
\end{split}\\
\begin{split} \label{eq:g3S18L}
\SDCs{2}{g}{\COcSaL}{}&=
\frac{N_c}{(N_c^2-1)}\frac{1-z}{z},
\end{split}\\
\begin{split}
\SDCs{2}{g}{\COaPaT}{}&=
\frac{N_c}{3(N_c^2-1)}\frac{1-z}{z^2}
\big[z^3+3z^2-12 z +3(3z-4)\mylog{1-z}\big],
\end{split}\\
\begin{split}
\SDCs{2}{g}{\COcPaL}{}&=\frac{B_F}{C_F}\times\SDCs{2}{g}{\CScPaL}{},
\end{split}\\
\begin{split}
\SDCs{2}{g}{\state{3}{P}{2, L \text{ or } T}{8}}{}&=\frac{B_F}{C_F}\times\SDCs{2}{g}{\state{3}{P}{2, L \text{ or } T}{8}}{},
\end{split}
\end{align}

\subsection{Different quark FFs}
\begin{equation}\label{eq:qto3S18L}
\SDCs{2}{q}{\COcSaL}{}=\frac{2(1-z)^2}{N_c z(\eta\, z^2-4z+4)},
\end{equation}
where $\eta=m_q^2/m_Q^2$.

\subsection{Same quark FFs}
\begin{align}
\begin{split}\label{eq:Qto3S11L}
\SDCs{2}{Q}{\CScSaL}{}&=
\frac{(N_c^2-1)^2}{6N_c^3}\frac{z(1-z)^2}{(z-2)^6}
(3z^4-24z^3+64z^2-32z+16),
\end{split}
\\
\SDCs{2}{Q}{\CSaPaL}{}&=
\frac{(N_c^2-1)^2}{30N_c^3}\frac{z(1-z)^2}{(z-2)^8}
(55z^6-232z^5+236z^4+224z^3\nonumber\\
&\hspace{1cm}
+592z^2-640z+320),
\\
\SDCs{2}{Q}{\CScPaL}{}&=
\frac{(N_c^2-1)^2}{15N_c^3}\frac{z(1-z)^2}{(z-2)^8}
(35z^6-312z^5+1136z^4\nonumber\\
&\hspace{1cm}
-2016z^3+1872z^2-960z+320),
\\
\SDCs{2}{Q}{\CScPbTa}{}&=
\frac{(N_c^2-1)^2}{15 N_c^3}\frac{z(1-z)^2}{(z-2)^8}
(75 z^6 -580 z^5 + 1628 z^4 \nonumber\\
&\hspace{1cm}
-1872 z^3 +1328 z^2 -512 z +128),
\\
\SDCs{2}{Q}{\CScPbTb}{}&=
\frac{8(N_c^2-1)^2}{15N_c^3}\frac{z(1-z)^4}{(z-2)^8}
(5z^4-32z^3+68z^2-32z+16),
\\
\begin{split}\label{eq:Qto3S18L}
\SDCs{2}{Q}{\COcSaL}{}&=
\frac{1}{6N_c^3}\frac{(1-z)^2}{z(z-2)^6}
\Big[12N_c^2(z-2)^4-12N_c z^2(z-4)(z-2)^2\\
&\hspace{1cm}
+z^2(3z^4-24z^3+64z^2-32z+16)\Big],
\end{split}\\
\SDCs{2}{Q}{\COaPaL}{}&=\frac{1}{(N_c^2-1)^2}\SDCs{2}{Q}{\CSaPaL}{},\\
\SDCs{2}{Q}{\COcPaL}{}&=\frac{1}{(N_c^2-1)^2}\SDCs{2}{Q}{\CScPaL}{},\\
\SDCs{2}{Q}{\state{3}{P}{2,T \text{ or } TT}{8}}{}&=\frac{1}{(N_c^2-1)^2}\SDCs{2}{Q}{\state{3}{P}{2, T \text{ or } TT}{8}}{}.
\end{align}

\subsection{$\statenc{3}{P}{J}$ operators with orbital angular momentum summed}

Here, we list the SDCs corresponding to the NRQCD LDMEs $\state{3,\lambda}{P}{}{1 \text{ or } 8}$ defined in Eq.~(\ref{eq:jpsi3pjapp}).
\begin{align}
\SDCs{2}{g}{\CScPL}{}&=
-\frac{1}{2N_c z^2}\Big[2(z-1)(z^2+8z-12)\mylog{1-z}\nonumber\\
&\hspace{1cm}
+z(2z^3+z^2-28z+24)\Big],\\
\SDCs{2}{g}{\COcPL}{}&=\frac{B_F}{C_F}\times\SDCs{2}{g}{\CScPL}{} \label{eq:gto3Pj8Polarized}\\
\SDCs{2}{Q}{\CScPL}{}&=
\frac{(N_c^2-1)^2}{6N_c^3}\frac{z(1-z)^2}{(z-2)^8}
(23z^6-192z^5+676z^4-1120z^3\nonumber\\
&\hspace{1cm}
+1104z^2-512z+192),
\\
\SDCs{2}{Q}{\COcPL}{}&=\frac{1}{(N_c^2-1)^2}\SDCs{2}{Q}{\CScPL}{}.
\end{align}

\subsection{Comparing with other calculations in the literature}\label{appsubsec:compare}

Polarized FFs in $g\to Q\bar{Q}(\COcSaL)$ channel has been calculated in Refs.~\cite{Beneke:1995yb,Braaten:2000pc}, our result in Eq.~(\ref{eq:g3S18L}) confirms their results.

FFs from $g\to Q\bar{Q}(\CScPa)$ and $g\to Q\bar{Q}(\CScPb)$ channels in different polarization states have been calculated in Ref.~\cite{Cho:1994gb} with cutoff regularization scheme. Comparing their results with ours in Eqs.~(\ref{eq:gto3P11L}), (\ref{eq:gto3P21L}) and (\ref{eq:gto3P21T}), we find that if we change $\log(\mu_\Lambda^2/m_Q^2)$ by $\log(\mu_\Lambda^2/m_Q^2)+5/3$ in our results, and take into account the different normalization of the LDMEs, we can reproduce their results in Ref.~\cite{Cho:1994gb}. The $5/3$ difference is due to different regularization scheme and was also realized in the calculation of polarization-summed FFs by authors of Ref.~\cite{Braaten:1996rp}. Therefore, our results of FFs from $g\to Q\bar{Q}(\CScPa)$ and $g\to Q\bar{Q}(\CScPb)$ channels in different polarization states are consistent with the results in Ref.~\cite{Cho:1994gb}.

Polarized FFs in $q\to Q\bar{Q}(\COcSaL)$, $Q\to Q\bar{Q}(\CScSaL)$ and $Q\to Q\bar{Q}(\COcSaL)$ channels have been reported in Ref.~\cite{Bodwin:2014bia} when we are preparing our paper.\footnote{Our results were published previously in Ref.~\cite{HongThesis}, in which there was a typo for $q \to Q\bar{Q}(\COcSaL) $ channel.} The authors of Ref.~\cite{Bodwin:2014bia} present their results with NRQCD LDMEs defined slightly different from ours. After taking into account the difference of the LDMEs, our results in Eqs.~(\ref{eq:qto3S18L}), (\ref{eq:Qto3S11L}) and (\ref{eq:Qto3S18L}) are consistent with theirs.

\section{Double-Parton Fragmentation Functions to a Polarized Heavy Quarkonium}
\label{app:DoublePolarized}

In this appendix we list the SDCs, up to order $O(\alpha_s)$, for all heavy quark pair fragmentation functions to a polarized $\cc$-pair in $S$-wave and $P$-wave, as defined in Eq.~(\ref{eq:NRQCDFacDouble}). The polarized FFs and unpolarized FFs are related by Eq.~(\ref{eq:PoUnPoRelation}). Therefore for outgoing $\cc$ with $n+1$ polarizations, we only give $n$ polarized FFs below. The other one can be derived by using Eq.~(\ref{eq:PoUnPoRelation}) and the unpolarized FFs calculated in Refs.~\cite{Ma:2013yla,Ma:2014eja}. All  channels that vanish all polarizations are not listed.

\subsection{$\Delta$-functions}
In our results below, we use the same definitions of $\Delta$-functions as that introduced in Ref.~\cite{Ma:2014eja}. We repeat these definitions below for readers' convenience and the completeness,
\begin{align}
\begin{split}
\DeltaZero
=
4\,\delta(\zeta_1)\delta(\zeta_2),
\end{split}\\
\begin{split}
\DeltaPPZero
=
4 \,z^2\,\delta'(\zeta_1)\delta'(\zeta_2),
\end{split}\\
\begin{split}
\DPMOne
=4\left[\DAA \pm \DBB\right] \left[\DXX \pm \DYY\right],
\end{split}\\
\begin{split}
&\DPMPOne
=-4\,z\Big\{
\left[\DPAA \pm \DPBB\right] \left[\DXX \pm \DYY\right]
\\
&\hspace{1cm}
+\left[\DAA\pm\DBB\right] \left[\DPXX\pm\DPYY\right]
\Big\},
\end{split}\\
\begin{split}
\DPMPPOne
=4\,z^2\,\left[\DPAA \pm \DPBB\right] \left[\DPXX \pm \DPYY\right],
\end{split}\\
\begin{split}
&\DPMEight
=4\Big\{(N_c^2-2)\left[\DAA\DXX+\DBB\DYY\right]\\
&\hspace{1cm}
\mp 2\left[\DAA\DYY+\DBB\DXX\right] \Big\},
\end{split}\\
\begin{split}
&\DPMPEight
= -4\, z\,\Big\{(N_c^2-2)\big[\DPAA\DXX+\DAA\DPXX
\\
&\hspace{1cm}
+\DPBB\DYY+\DBB\DPYY\big]
\\
&\hspace{1cm}
\mp 2\big[\DPAA\DYY+\DAA\DPYY
\\
&\hspace{1cm}
+\DPBB\DXX+\DBB\DPXX\big]
\Big\},
\end{split}\\
\begin{split}
&\DPMPPEight
=4\,z^2\,\Big\{(N_c^2-2)\left[\DPAA\DPXX+\DPBB\DPYY\right]\\
&\hspace{1cm}
\mp 2\left[\DPAA\DPYY+\DPBB\DPXX\right] \Big\},
\end{split}
\end{align}
All these $\Delta$-functions are invariant under the transformation ($\zeta_1 \to -\zeta_1$, $\zeta_2 \to -\zeta_2$) and the exchange $\zeta_1 \leftrightarrow \zeta_2$, including the crossing exchange
($\zeta_1 \to -\zeta_2$, $\zeta_2 \to -\zeta_1$).
In addition, $\DeltaZero$, $\DPlusOne$, $\DPlusPOne$ and $\DPlusPPOne$ are even in both $\zeta_1$ and $\zeta_2$, while $\DeltaPPZero$, $\DMinusOne$, $\DMinusPOne$ and $\DMinusPPOne$ are odd in both $\zeta_1$ and $\zeta_2$.
Under the integration of $\zeta_1$ and $\zeta_2$ with a well behaved test function, the asymptotic behaviors of these $\Delta$-functions at $z\to1$ are
%
\begin{align}
\begin{split}
\lim_{z \to 1}\DPlusOne =O[1],
\hspace{3.15cm}
\lim_{z \to 1}\DMinusOne =O[(1-z)^2],
\end{split}
\nonumber
\\
\begin{split}
\lim_{z \to 1}\DPlusPOne =O[(1-z)],
\hspace{2cm}
\lim_{z \to 1}\DMinusPOne =O[(1-z)],
\end{split}
\nonumber
\\
\begin{split}
\lim_{z \to 1}\DPlusPPOne =O[(1-z)^2],
\hspace{1.75cm}
\lim_{z \to 1}\DMinusPPOne =O[1],
\end{split}
\nonumber
\\
\begin{split}
\lim_{z \to 1}\DPMEight =O[1],
\hspace{3.05cm}
\lim_{z \to 1}\DPMPEight =O[(1-z)],
\end{split}
\nonumber
\\
\begin{split}
\lim_{z \to 1}\DPMPPEight =O[1],
\end{split}
\nonumber
\end{align}
Therefore,
\begin{align}
\begin{split}
\frac{\DMinusOne}{(1-z)},
\hspace{0.3cm}
\frac{\DPMPOne}{(1-z)},
\hspace{0.3cm}
\frac{\DPlusPPOne}{(1-z)},
\hspace{0.3cm}
\text{and}
\hspace{0.2cm}
\frac{\DPMPEight}{(1-z)}\nonumber
\end{split}
\end{align}
do not exhibit any pole at $z=1$.

\subsection{Leading Order}

\begin{align}
\SDC{0}{\vone}{\CScSaT}{}&=0,\\
\SDC{0}{\vone}{\CScPbTb}{}&=0,\\
\SDC{0}{\vone}{\CScPbTa}{}&=0,\\
\SDC{0}{\aone}{\CSaPaT}{}&=0,\\
\SDC{0}{\aone}{\CScPaT}{}&=0,\\
\SDC{0}{\tone}{\CScSaL}{}&=0,\\
\SDC{0}{\tone}{\CSaPaL}{}&=0,\\
\SDC{0}{\tone}{\CScPaL}{}&=0,\\
\SDC{0}{\tone}{\CScPbTb}{}&=0,\\
\SDC{0}{\tone}{\CScPbL}{}&=0.
\end{align}
The corresponding color octet channels also vanish.

\subsection{NLO - Vector}
\begin{align}
\SDC{1}{\vone}{\CScSaT}{}&=0,\\
\SDC{1}{\vone}{\CScPbTa}{}&=0,\\
\SDC{1}{\vone}{\CScPbTb}{}&=0,\\
\SDC{1}{\vone}{\COcSaT}{}&=\frac{1}{8 N_c}(1-z)z\DMinusOne,\\
\SDC{1}{\vone}{\COaPaT}{}&=\frac{1}{12 N_c}(1-z)z\DMinusOne,\\
\SDC{1}{\vone}{\COcPaT}{}&=\frac{1}{96 N_c}
\Big\{
8\DeltaZero \delta(1-z)
\Big(-\LogIR+2\logtwo+\frac{1}{2}\Big)\nonumber\\
&\hspace{0cm}
+6\DPlusPPOne\,z\,(1-z)
-3\DPlusPOne\,z\,(2z-3)\nonumber\\
&\hspace{0cm}
+4\,z\, \DPlusOne
\Big[\frac{1}{(1-z)_+}-2z+3\Big]
\Big\},\\
\SDC{1}{\vone}{\COcPbTa}{}&=\frac{1}{96 N_c}
\Big\{
8\DeltaZero \delta(1-z)
\Big(-\LogIR+2\logtwo+\frac{1}{2}\Big)\nonumber\\
&\hspace{0cm}
+6\DPlusPPOne\,z\,(1-z)
-3\DPlusPOne\,z\,(2z-1)\nonumber\\
&\hspace{0cm}
+4\,z\, \DPlusOne
\Big[\frac{1}{(1-z)_+}-2z+1\Big]
\Big\},\\
\SDC{1}{\vone}{\COcPbTb}{}&=\frac{1}{24 N_c}z(1-z)\DPlusOne,\\
\SDC{1}{\veight}{\COcSaT}{}&=\frac{1}{8 N_c(N_c^2-1)}(1-z)z\DPlusEight,\\
\SDC{1}{\veight}{\COaPaT}{}&=\frac{1}{12 N_c(N_c^2-1)}(1-z)z\DPlusEight,\\
\SDC{1}{\veight}{\COcPaT}{}&=\frac{1}{96 N_c(N_c^2-1)}
\Big\{
4(N_c^2-4)\DeltaZero \delta(1-z)
\Big(-\LogIR\nonumber\\
&\hspace{0cm}
+2\logtwo+\frac{1}{2}\Big)
+6\DMinusPPEight\,z\,(1-z)
-3\DMinusPEight\,z\,(2z-3)\nonumber\\
&\hspace{0cm}
+4\,z\, \DMinusEight
\Big[\frac{1}{(1-z)_+}-2z+3\Big]
\Big\},\\
\SDC{1}{\veight}{\COcPbTa}{}&=\frac{1}{96 N_c(N_c^2-1)}
\Big\{
4(N_c^2-4)\DeltaZero \delta(1-z)
\Big(-\LogIR\nonumber\\
&\hspace{0cm}
+2\logtwo+\frac{1}{2}\Big)
+6\DMinusPPEight\,z\,(1-z)
-3\DMinusPEight\,z\,(2z-1)\nonumber\\
&\hspace{0cm}
+4\,z\, \DMinusEight
\Big[\frac{1}{(1-z)_+}-2z+1\Big]
\Big\},\\
\SDC{1}{\veight}{\COcPbTb}{}&=\frac{1}{24 N_c(N_c^2-1)}z(1-z)\DMinusEight,\\
\SDC{1}{\veight}{\state{{2S+1}}{L}{J,\lambda}{1}}{}&=\SDC{1}{\vone}{\state{{2S+1}}{L}{J,\lambda}{8}}{},
\end{align}
where $\lambda$ is the polarization of the outgoing heavy quark pair.

\subsection{NLO - Axial-vector}
\begin{align}
\SDC{1}{\aone}{\CSaPaT}{}&=0,\\
\SDC{1}{\aone}{\CScPaT}{}&=0,\\
\SDC{1}{\aone}{\COcSaT}{}&=\frac{1}{8 N_c}z(1-z)\DPlusOne,\\
\SDC{1}{\aone}{\COaPaT}{}&=\frac{1}{24 N_c}
\Big\{
4\DeltaZero \delta(1-z)
\Big(-\LogIR+2\logtwo+\frac{1}{2}\Big)\nonumber\\
&\hspace{0cm}
+2\,z\, \DPlusOne
\Big[\frac{1}{(1-z)_+}-\frac{3}{2}z\Big]
\Big\},\\
\SDC{1}{\aone}{\COcPaT}{}&=\frac{1}{96 N_c}
z(1-z)\big(6\DMinusPPOne+9\DMinusPOne+16\DMinusOne),\\
\SDC{1}{\aone}{\COcPbTa}{}&=\frac{1}{96 N_c}
z(1-z)\big(6\DMinusPPOne+3\DMinusPOne+8\DMinusOne),\\
\SDC{1}{\aone}{\COcPbTb}{}&=\frac{1}{24 N_c}z(1-z)\DMinusOne,\\
\SDC{1}{\aeight}{\COcSaT}{}&=\frac{1}{8 N_c(N_c^2-1)}z(1-z)\DMinusEight,\\
\SDC{1}{\aeight}{\COaPaT}{}&=\frac{1}{24 N_c(N_c^2-1)}
\Big\{
2(N_c^2-4)\DeltaZero \delta(1-z)
\Big(-\LogIR\nonumber\\
&\hspace{0cm}
+2\logtwo+\frac{1}{2}\Big)
+2\,z\, \DMinusEight
\Big[\frac{1}{(1-z)_+}-\frac{3}{2}z\Big]
\Big\},\\
\SDC{1}{\aeight}{\COcPaT}{}&=\frac{1}{96 N_c(N_c^2-1)}
z(1-z)\big(6\DPlusPPEight+9\DPlusPEight+16\DPlusEight),\\
\SDC{1}{\aone}{\COcPbTa}{}&=\frac{1}{96 N_c(N_c^2-1)}
z(1-z)\big(6\DPlusPPEight+3\DPlusPEight+8\DPlusEight),\\
\SDC{1}{\aone}{\COcPbTb}{}&=\frac{1}{24 N_c(N_c^2-1)}z(1-z)\DPlusEight,\\
\SDC{1}{\aeight}{\state{{2S+1}}{L}{J,\lambda}{1}}{}&=\SDC{1}{\aone}{\state{{2S+1}}{L}{J,\lambda}{8}}{},
\end{align}
where $\lambda$ is the polarization of the outgoing heavy quark pair.

\subsection{NLO - Tensor}
\begin{align}
\SDC{1}{\tone}{\CScSaL}{}&=0,\\
\SDC{1}{\tone}{\CSaPaL}{}&=0,\\
\SDC{1}{\tone}{\CScPaL}{}&=0,\\
\SDC{1}{\tone}{\CScPbTb}{}&=0,\\
\SDC{1}{\tone}{\CScPbL}{}&=0,\\
\SDC{1}{\tone}{\COcSaL}{}&=\frac{1}{16N_c}z(1-z)\DMinusOne,\\
\SDC{1}{\tone}{\COaPaL}{}&=\frac{1}{48N_c}z(1-z)\DMinusOne,\\
\SDC{1}{\tone}{\COcPaL}{}&=
\frac{1}{48 N_c}
\Big\{
2\DeltaZero \delta(1-z)
\Big(-\LogIR+2\logtwo+\frac{1}{2}\Big)\nonumber\\
&\hspace{0cm}
+z\, \DPlusOne
\Big[\frac{1}{(1-z)_+}-4z+1\Big]
\Big\},\\
\SDC{1}{\tone}{\COcPbTb}{}&=
\frac{1}{24 N_c}
\Big\{
2\DeltaZero \delta(1-z)
\Big(-\LogIR+2\logtwo+\frac{1}{2}\Big)\nonumber\\
&\hspace{0cm}
+z\, \DPlusOne
\Big[\frac{1}{(1-z)_+}-z+1\Big]
\Big\},\\
\SDC{1}{\tone}{\COcPbL}{}&=\frac{1}{288 N_c}
\Big\{
4\DeltaZero \delta(1-z)
\Big(-\LogIR+2\logtwo+\frac{1}{2}\Big)\nonumber\\
&\hspace{0cm}
+12\,z\DPlusPPOne\,(1-z)
-3\,z\DPlusPOne\,(3z-2)\nonumber\\
&\hspace{0cm}
+2\,z\, \DPlusOne
\Big[\frac{1}{(1-z)_+}-7z+5\Big]
\Big\},\\
\SDC{1}{\teight}{\COcSaL}{}&=\frac{1}{16N_c(N_c^2-1)}z(1-z)\DPlusEight,\\
\SDC{1}{\teight}{\COaPaL}{}&=\frac{1}{48N_c(N_c^2-1)}z(1-z)\DPlusEight,\\
\SDC{1}{\teight}{\COcPaL}{}&=
\frac{1}{48 N_c (N_c^2-1)}
\Big\{
(N_c^2-4)\DeltaZero \delta(1-z)
\Big(-\LogIR\nonumber\\
&\hspace{0cm}
+2\logtwo+\frac{1}{2}\Big)
+z\, \DMinusEight
\Big[\frac{1}{(1-z)_+}-4z+1\Big]
\Big\},\\
\SDC{1}{\teight}{\COcPbTb}{}&=
\frac{1}{24 N_c (N_c^2-1)}
\Big\{
(N_c^2-4)\DeltaZero \delta(1-z)
\Big(-\LogIR\nonumber\\
&\hspace{0cm}
+2\logtwo+\frac{1}{2}\Big)
+z\, \DMinusEight
\Big[\frac{1}{(1-z)_+}-z+1\Big]
\Big\},\\
\SDC{1}{\teight}{\COcPbL}{}&=\frac{1}{288 N_c(N_c^2-1)}
\Big\{
2(N_c^2-4)\DeltaZero \delta(1-z)
\Big(-\LogIR\nonumber\\
&\hspace{0cm}
+2\logtwo+\frac{1}{2}\Big)
+12\,z\DMinusPPEight\,(1-z)
-3\,z\DMinusPEight\,(3z-2)\nonumber\\
&\hspace{0cm}
+2\,z\, \DMinusEight
\Big[\frac{1}{(1-z)_+}-7z+5\Big]
\Big\},\\
\SDC{1}{\teight}{\state{{2S+1}}{L}{J,\lambda}{1}}{}&=\SDC{1}{\tone}{\state{{2S+1}}{L}{J,\lambda}{8}}{},
\end{align}
where $\lambda$ is the polarization of the outgoing heavy quark pair.


\subsection{$\statenc{3}{P}{J}$ operators with orbital angular momentum summed}


In this subsection, we list all SDCs corresponding to the NRQCD LDMEs $\state{3,\lambda}{P}{}{1 \text{ or } 8}$ defined in Eq.~(\ref{eq:jpsi3pjapp}).

\subsubsection{Leading Order}
\begin{align}
\SDC{0}{\vone}{\CScPT}{}&=0,\\
\SDC{0}{\aone}{\CScPL}{}&=0,\\
\SDC{0}{\tone}{\CScPL}{}&=0.
\end{align}
The corresponding color octet channels also vanish.

\subsubsection{Next-to-leading order}

\begin{align}
\SDC{1}{\vone}{\CScPT}{}&=0,\\
\SDC{1}{\aone}{\CScPL}{}&=0,\\
\SDC{1}{\tone}{\CScPL}{}&=0,\\
\SDC{1}{\vone}{\COcPT}{}&=\frac{1}{8 N_c}z(1-z)
\Big\{\DPlusOne\LogUV
-\DPlusOne\Big[2\mylog{2(1-z)}-1\Big]\nonumber\\
&\hspace{0.5cm}
+\DPlusPOne+\DPlusPPOne\Big\},
\\
\SDC{1}{\veight}{\CScPT}{}&=\SDC{1}{\vone}{\COcPT}{},
\\
\SDC{1}{\veight}{\COcPT}{}&=\frac{1}{8 N_c(N_c^2-1)}z(1-z)
\Big\{\DMinusEight\LogUV\nonumber\\
&\hspace{0.5cm}
-\DMinusEight\Big[2\mylog{2(1-z)}-1\Big]
+\DMinusPEight+\DMinusPPEight\Big\},
\\
\SDC{1}{\aone}{\COcPL}{}&=\frac{1}{16N_c}z(1-z)
\Big\{
(\DMinusOne+\DMinusPOne+\DMinusPPOne)\Big[\LogUV \nonumber\\
&\hspace{0.5cm}
-2\mylog{2(1-z)}-3\Big]-\DMinusPOne
\Big\},\\
\SDC{1}{\aeight}{\CScPL}{}&=\SDC{1}{\aone}{\COcPL}{},
\\
\SDC{1}{\aeight}{\COcPL}{}&=\frac{1}{16N_c(N_c^2-1)}z(1-z)
\Big\{
(\DPlusEight+\DPlusPEight+\DPlusPPEight)\Big[\LogUV \nonumber\\
&\hspace{0.5cm}
-2\mylog{2(1-z)}-3\Big]-\DPlusPEight
\Big\}
,\\
\SDC{1}{\tone}{\COcPL}{}&=\frac{1}{16N_c}z(1-z)(2\DPlusOne+\DPlusPOne+\DPlusPPOne),\\
\SDC{1}{\teight}{\CScPL}{}&=\SDC{1}{\tone}{\COcPL}{},\\
\SDC{1}{\teight}{\COcPL}{}&=\frac{1}{16N_c(N_c^2-1)}z(1-z)(2\DMinusEight+\DMinusPEight+\DMinusPPEight).
\end{align}

\providecommand{\href}[2]{#2}\begingroup\raggedright
\endgroup

\end{document}